\documentclass[fleqn,usenatbib]{mnras}

\usepackage[flushleft]{threeparttable}
\usepackage{longtable}  
\DeclareUnicodeCharacter{2212}{\textendash}
\usepackage{epstopdf}

\usepackage{mathptmx}

\usepackage[T1]{fontenc}

\DeclareRobustCommand{\VAN}[3]{#2}
\let\VANthebibliography\thebibliography
\def\thebibliography{\DeclareRobustCommand{\VAN}[3]{##3}\VANthebibliography}

\usepackage{graphicx}
\usepackage{amsmath}

\usepackage{wrapfig}
\usepackage{lscape}
\usepackage{rotating}
\usepackage[none]{hyphenat}

\usepackage{txfonts}

\newcommand{\rbar}{$R_{\rm bar}$}

\newcommand{\omegabar}{$\Omega_{\rm bar}$}
\newcommand{\sbar}{$S_{\rm bar}$}

\newcommand{\rr}{${\cal R}$}
\newcommand{\vcirc}{$V_{\rm circ}$}
\newcommand{\msun}{\rm M_{\odot}}
\newcommand{\rpetro}{$R_{\rm Petro}$}

\newcommand{\rmean}{$R_{\rm bar}$}
\newcommand{\rcor}{$R_{\rm cor}$}
\newcommand{\rpha}{$R_{\phi_2}$}

\title[Bars in cluster dwarf galaxies]{The dynamical state of bars in cluster dwarf galaxies: \\The cases of NGC~4483 and NGC~4516}

\author[V. Cuomo et al.]{Virginia Cuomo,$^{1}$\thanks{E-mail: virginia.cuomo@uda.cl}
Lorenzo Morelli,$^{1}$
J. Alfonso L. Aguerri,$^{2,3}$
Enrico Maria Corsini,$^{4,5}$
\newauthor Victor P. Debattista,$^{6}$
Lodovico Coccato,$^{7}$ 
Alessandro Pizzella,$^{4,5}$
Alessandro Boselli,$^{8}$
Chiara Buttitta,$^{9}$
\newauthor Adriana de Lorenzo-Cáceres,$^{2,3}$
Laura Ferrarese,$^{10}$ 
Daniele Gasparri,$^{1}$
Yun Hee Lee,$^{11,12}$ 
\newauthor Jairo Mendez-Abreu,$^{2,3}$
Joel Roediger,$^{10}$ 
and Stefano Zarattini$^{2,3}$ \\
$^{1}$Instituto de Astronomía y Ciencias Planetarias, Universidad de Atacama, Avenida Copayapu 485, 1530000 Copiapó, Atacama, Chile\\
$^{2}$Instituto de Astrofísica de Canarias, calle Vía Láctea s/n, 38205 La Laguna, Tenerife, Spain\\
$^{3}$Departamento de Astrofísica, Universidad de La Laguna, Avenida Astrofísico Francisco Sánchez s/n, 38206 La Laguna, Tenerife, Spain \\
$^{4}$ Dipartimento di Fisica e Astronomia “G. Galilei”, Università di Padova, vicolo dell’Osservatorio 3, I-35122 Padova, Italy \\
$^{5}$ INAF-Osservatorio Astronomico di Padova, vicolo dell’Osservatorio 2, I-35122 Padova, Italy \\
$^{6}$ Jeremiah Horrocks Institute, University of Central Lancashire,PR1 2HE  Preston, UK \\
$^{7}$ European Southern Observatory, Karl-Schwarzschild-Strasse 2, D-85748 Garching, Germany \\
$^{8}$ Aix Marseille Univ, CNRS, CNES, LAM, Marseille, France \\
$^{9}$ INAF - Osservatorio Astronomico di Capodimonte, Salita Moiariello 16, I-80131 Napoli, Italy\\
$^{10}$ National Research Council of Canada, Herzberg Astronomy and Astrophysics Research Centre, 5071 West Saanich Road, BC V9E 2E7 Victoria, Canada\\
$^{11}$ Department of Astronomy and Atmospheric Sciences, Kyungpook National University, Daegu 41566, Republic of Korea \\
$^{12}$ Korea Astronomy and Space Science Institute (KASI), 776 Daedeokdae-ro, Yuseong-gu, Daejeon 34055, Republic of Korea \\
 }

\date{Accepted XXX. Received YYY; in original form ZZZ}

\pubyear{2023}

\begin{document}
\label{firstpage}
\pagerange{\pageref{firstpage}--\pageref{lastpage}}
\maketitle

\begin{abstract}
Dwarf barred galaxies are the perfect candidates for hosting slowly-rotating bars. They are common in dense environments and they have a relatively shallow potential well, making them prone to heating by interactions. When an interaction induces bar formation, the bar should rotate slowly. They reside in massive and centrally-concentrated dark matter halos, which slow down the bar rotation through dynamical friction. While predictions suggest that slow bars should be common, measurements of bar pattern speed, using the Tremaine-Weinberg method, show that bars are mostly fast in the local Universe. We present a photometric and kinematic characterisation of bars hosted by two dwarf galaxies in the Virgo Cluster, NGC~4483 and NGC~4516. We derive the bar length and strength using the Next Generation Virgo Survey imaging and the circular velocity, bar pattern speed, and rotation rate using spectroscopy from the Multi Unit Spectroscopic Explorer. Including the previously studied galaxy IC~3167, we compare the bar properties of the three dwarf galaxies with those of their massive counterparts from literature. Bars in the dwarf galaxies are shorter and weaker, and rotate slightly slower with respect to those in massive galaxies. This could be due to a different bar formation mechanism and/or to a large dark matter fraction in the centre of dwarf galaxies. We show that it is possible to push the application of the Tremaine-Weinberg method to the galaxy low mass regime. 
\end{abstract}




\section{Introduction}

Stellar bars are hosted in about 70 per cent of nearby massive disc galaxies \citep{buta2015,DiazGarcia2016,erwin2018}, including the Milky Way \citep{BlandHawthorn2016}. Stars in a bar follow elongated orbits aligned with the bar major-axis, while the bar rotates as a coherent structure around the galaxy centre with a fixed angular frequency, namely the bar pattern speed, \omegabar\ \citep[e.g.][]{Binney2008,Cuomo2020}. The rotation of the bar drives the redistribution of mass and angular momentum in the disc. Therefore, both the morphology and dynamics of a barred galaxy depend on \omegabar\ \citep[e.g.][]{athanassoula2003,Petersen2019}.

The bar pattern speeed \omegabar\ is usually parametrised by the bar rotation rate, \rr~ = \rcor/\rbar, the distance-independent ratio between the corotation radius, \rcor, and bar length, \rbar, which marks the extension of the stellar orbits supporting the bar. The corotation radius is linked to \omegabar, since \rcor~ = \vcirc/\omegabar,
where \vcirc\ is the circular velocity of the galaxy. Bars are classified as fast if $1.0 \leq {\cal R} \leq 1.4$ (i.e. when they extend out to \rcor\ and rotate as fast as possible), and as slow if \rr~$> 1.4$ (i.e. when they fall short of \rcor\ and have a slower rotation; \citealt{Athanassoula1992,Debattista2000}). 
Self-consistent bars cannot extend beyond the corotation radius, due to elliptical periodic orbits being unstable
in those regions \citep{contopoulos1980}. Therefore, \rr\ is expected to be $> 1.0$.

Analytical work \citep{weinberg1985} and numerical simulations \citep{Debattista1998,Athanassoula2013} show that during galaxy evolution, a bar is slowed through the exchange of angular momentum with the other galaxy components, while the bar length and strength, \sbar\ (i.e. the contribution of the bar to the galaxy potential), tend to increase. This is particularly true when a massive and centrally-concentrated dark matter (DM) halo is present, since it means that more mass is available to absorb angular momentum near the resonances resulting in strong dynamical friction. \cite{Debattista2000} put constraints on the DM distribution in the inner regions of barred galaxies and argued that galaxies hosting fast bars should be embedded in DM halos with a low central density, contrary to the predictions of cosmologically-motivated simulations \citep[e.g.][]{navarro1996}. On the other hand, simulated bars induced by tidal interactions are born slow and stay slow for a long time during their evolution \citep{MartinezValpuesta2017}. The efficiency of their formation depends on the orbital configuration of the encounter \citep{Lokas2018} and on the properties of the galaxies involved \citep{Gadja2018}. In fact, the formation and evolution of a barred galaxy is a multi-parameter problem \citep{athanassoula2003,Athanassoula2013}. Therefore, the characterisation of the bar properties, and in particular the measurement of \rr, is desirable both to investigate the secular evolution of barred galaxies and to test whether the measured DM distribution matches the predictions of cold DM cosmology.

The Tremaine-Weinberg method \citep[hereafter TW]{Tremaine1984} is a model-independent way to measure \omegabar. It requires knowing the surface luminosity density and line-of-sight (LOS) velocity distribution of a tracer satisfying the continuity equation \citep[see][]{Borodina2023}.

This technique has been largely applied to galaxies in the local Universe and compared to gas based methods, finding similar results \citep[see e.g.][]{Rautiainen2008,Beckman2018}. Indeed, more than 300 galaxies have been analysed so far with the TW method, using both long-slit spectroscopy \citep{Corsini2011}, and integral-field spectroscopy (IFS) from both dedicated observations \citep{Debattista2004,Cuomo2019,Buttitta2022} and publicly available data from large surveys \citep{Aguerri2015,Cuomo2019b,Guo2019,Williams2021,Geron2023}. Neglecting cases with large uncertainties and bars within the unstable regime (${\cal R}< 1.0$), nearly all the analysed bars are consistent with being fast: their host galaxies are not dominated by DM in the central regions and the formation of their bars is not driven by galaxy interactions. 


Dwarf galaxies are commonly thought to host a massive and centrally-concentrated DM halo \citep{Cote1991}, which may cause strong dynamical friction capable of slowing down the bar. Dwarf galaxies have small sizes (with effective radii $\lesssim 2$~kpc, e.g. \citealt{Eigenthaler2018}) and low stellar masses ($\lesssim 10^9~\msun$, e.g. \citealt{Eigenthaler2018,Michea2021}). They are the dominant galaxy population in clusters and groups \citep{Binney1987,Geha2012}, and present various kinematical and morphological characteristics, a wide range in rotation, and hidden disc features, such as bars and spiral structures \citep{Lisker2006,Rys2013,Michea2021}. In the Virgo Cluster, bars and lenses are hosted in about half of the dwarf galaxies with disc features \citep{Janz2014}. These findings have resulted in the definition of the class of disc dwarf galaxies. Their wide range in properties remains not completely understood but it has to be linked to the mechanisms involved in their formation and/or to environmental factors \citep{Lisker2013}. Simulations suggest that the formation of bars in dwarf galaxies is triggered by the cluster tidal field only near the cluster core \citep{Mastropietro2005,Lokas2014}. But, dwarf galaxies are intrinsically unstable to bar formation even without external forces, so they can be found everywhere in the cluster \citep{Barazza2002,Kwak2017}. However, dwarf barred galaxies with stellar mass $<10^9~\msun$ seems to be rarer than in the field, probably due to the fact that interactions in dense environments tend to heat the discs and suppress the bar instability \citep{MendezAbreu2010,mendezabreu2012}.

To date we know \omegabar\ and \rr\ for only two dwarf barred galaxies. The first one is NGC~4431, which hosts a fast bar, albeit with a large uncertainty, due to limitations of long-slit spectroscopy used to apply the TW method ($\Delta {\cal R}/{\cal R} > 1.3$; \citealt{Corsini2007}). The second one is IC~3167, recently analysed with IFS by \cite{Cuomo2022}, which hosts a slow lopsided bar. The peculiar shape of the bar together with its recovered dynamical properties suggest it could be formed through the interaction of the galaxy with the Virgo Cluster, into which it is falling. 

Here, we present the characterisation of the bars of NGC~4483 and NGC~4516, two dwarf barred galaxies in the Virgo Cluster. In particular, we derive \rbar\ and \sbar\ using deep photometry from the Next Generation Virgo Survey \citep[NGVS,][]{Ferrarese2012} and \omegabar\ from integral-field spectroscopy performed with the Multi Unit Spectroscopic Explorer \citep[MUSE,][]{bacon2010} of the European Southern Observatory. Including the results for IC~3167 from \cite{Cuomo2022}, we discuss the bar properties of the three dwarf galaxies and compare them with those of their more massive counterparts. 

We structure the paper as follows. We present the properties of NGC~4483 and NGC~4516 in Section~\ref{sec:sample_emvironment}. We show the broad-band imaging and photometric properties of the bars in Section~\ref{sec:images}, and the integral-field spectroscopy and kinematic and dynamical properties of the bars in Section~\ref{sec:muse}. We analyse the bar properties and compare them to those of bars in more massive galaxies in Section~\ref{sec:results}. We discuss our findings in Section~\ref{sec:discussion} and report our conclusions in Section~\ref{sec:conclusion}. Throughout the paper, we adopt as cosmological parameters $\Omega_{\rm m} = 0.286$, $\Omega_{\Lambda}= 0.714$, and $H_0 = 69.3$~km~s$^{-1}$~Mpc$^{-1}$ \citep{Hinshaw2013}.

\section{Properties of NGC~4483 and NGC~4516}
\label{sec:sample_emvironment}

We analyse two dwarf barred galaxies located in the Virgo Cluster \citep{Binggeli1985,kim2014}, for which we obtained IFS data with MUSE (Prog. Id.: 0106.B-0158(A), P.I.: V. Cuomo). Deep optical images of these galaxies have been obtained in the $u, g, i$ and $z$ filters as part of the NGVS. 

\begin{figure*}
    \centering
   \includegraphics[scale=0.35]{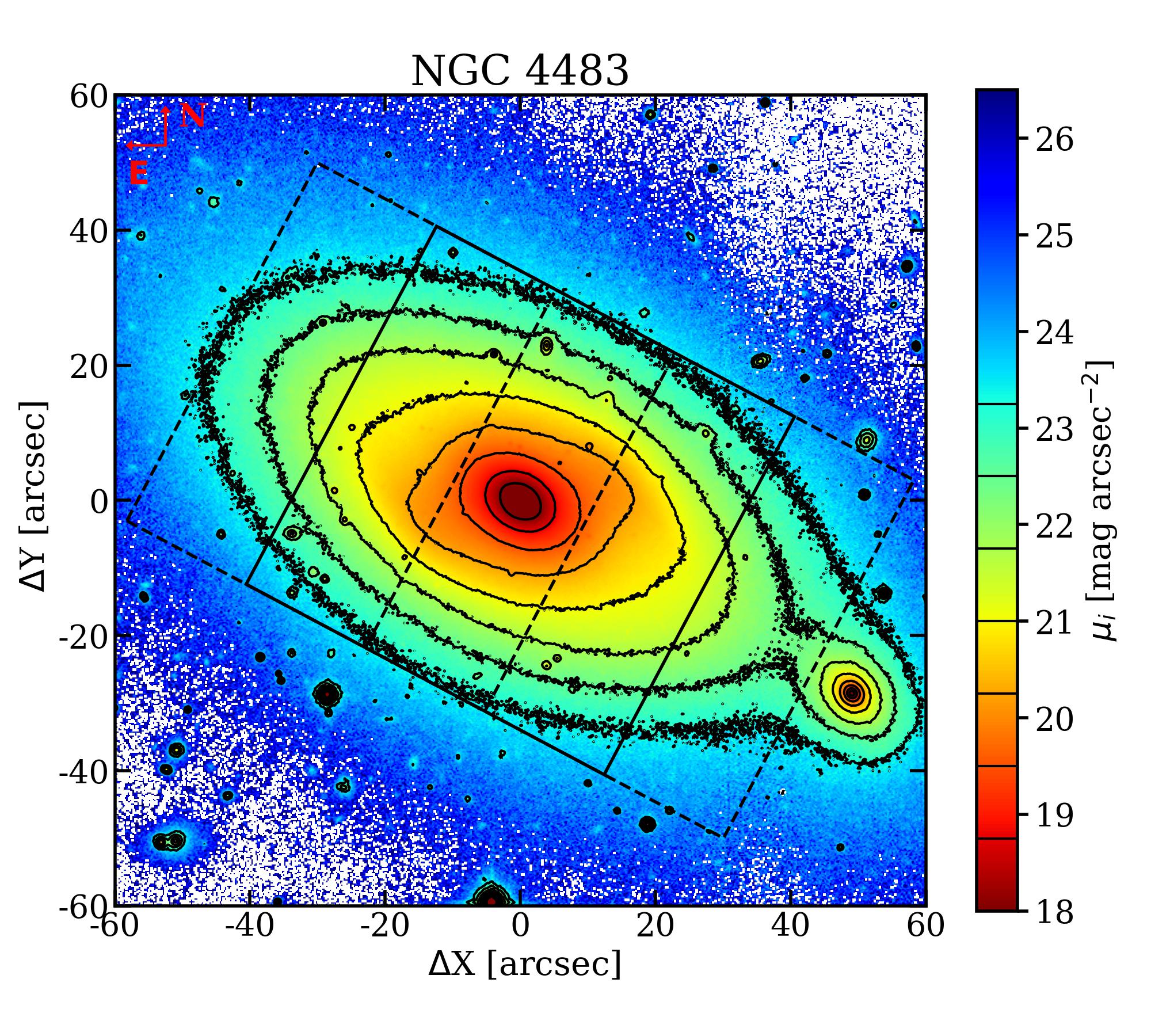}\vspace{-0.4cm}
    \includegraphics[scale=0.35]{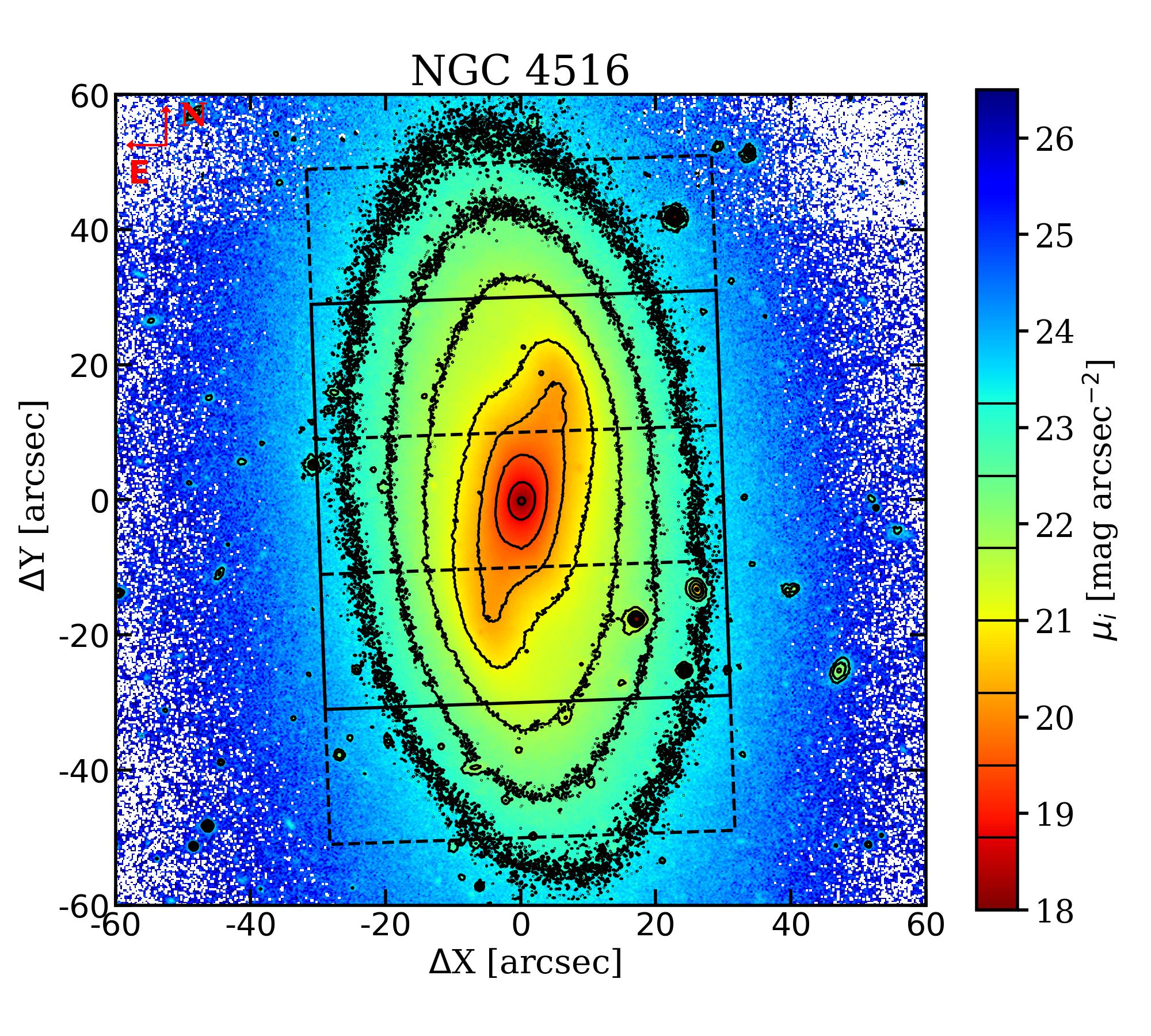}\vspace{-0.4cm}
   \caption{NGVS $i$-band images of the sample galaxies. The black contours mark some of the isophotes and the corresponding surface brightness levels are reported on the colourbar (black ticks). The black solid and dashed squares show the field of view of the MUSE pointings.}
   \label{fig:sample}
\end{figure*}

\begin{figure}
    \centering
    \includegraphics[scale=0.42]{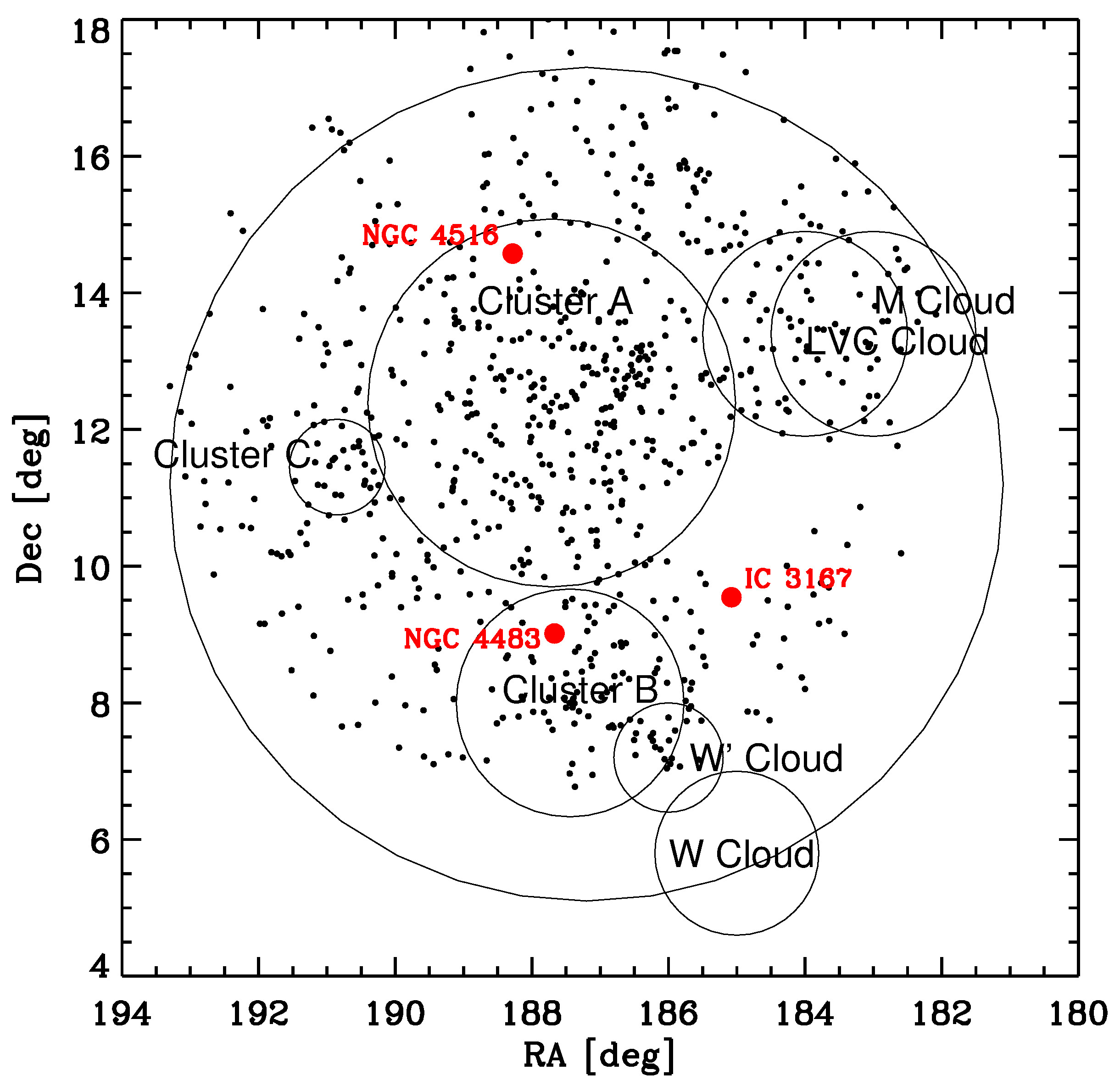}
    \caption{Location of NGC~4483 and NGC~4516 from the NGVS and of IC~3167 from Cuomo et al. (2022) within the Virgo Cluster. The different Virgo Cluster substructures are marked (Boselli et al. (2018).}
    \label{fig:galaxies_virgo}
\end{figure}

Figure~\ref{fig:sample} shows the NGVS $i$ band images of the sample, with superimposed the square footprint of the three MUSE pointings needed to cover the extension of the discs, while Table~\ref{tab:galaxies_poperties} reports the properties of the two galaxies analysed here and those of IC~3167. NGC~4483 and NGC~4516 are located at different distances within the Virgo Cluster (Fig.~\ref{fig:galaxies_virgo}) but both are in the central virialised region of the cluster. On the contrary, IC~3167 is located in the in-falling region, as discussed by \cite{Bidaran2020}.

NGC~4483 is an SB0 galaxy, with a stellar mass of ${M}_{\ast}=7.5\times10^9~\msun$ \citep{Roediger2017}, is located at an angular distance of 3.3\degr\ from M~87 \citep{Kashibadze2020}. It lives in the virial core of the Virgo Cluster. It is in an intermediate density environment ($\Sigma_5=0.45$~Mpc$^{-2}$; see also \citealt{deBortoli2022}) and its closest bright galaxy is IC~3430, located at a projected distance of 0.6~Mpc. NGC~4483 hosts a bar with an ansae morphology, associated with a barlens \citep{Buta2019}, and a nuclear stellar disc \citep{ledo2010}. It possibly hosts a boxy/peanut (BP)-shaped bulge \citep{Bureau1999b}.

NGC~4516 is an SBa galaxy (with ${M}_{\ast}=5.7\times10^9~\msun$ \citep{Roediger2017}, located in the virial core of the Virgo Cluster at an angular distance of 2.3\degr\ from M~87 \citep{Kashibadze2020}. It lives in a low density environment ($\Sigma_5=0.15$~Mpc$^{-2}$) and its closest bright galaxy is IC~3478, located at a projected distance of 2.1~Mpc. NGC~4516 clearly shows an X-shaped structure in the central region, which is associated with a BP bulge, and S-shaped spiral arms, which emerge from the ends of the bar \citep{Buta2019}. 

The three dwarf galaxies included in our sample are lenticular or early-type spiral galaxies. Early-type dwarf galaxies are quite common in dense environments, such as the central region of the Virgo cluster \citep[e.g.][]{Lisker2009}, and their morphological properties make them well suited for the application of the TW method \citep{Corsini2011}.

\begin{table*}
\begin{center}
\caption{\label{tab:galaxies_poperties} Properties of the sample of galaxies.}
\begin{tabular}{ccccccccc}
\hline
Galaxy  & Alt. name & NGVS name & Morph. type & $m_{r,i}$ & $M_{r,i}$ & $R_{50}$ & $M_{*}$ & $\Sigma_5$ \\
  & & &  & {[}mag{]} & {[}mag{]} & {[}arcsec{]} & {[}10$^9$ ${\rm M}_{\odot}${]} & [Mpc$^{-2}$] \\ 
(1) & (2) & (3) & (4) & (5) & (6) & (7) & (8) & (9) \\     
\hline
NGC~4483 & VCC1303 & NGVSJ12:30:40.64+09:00:56.4 & SB0 & 12.07,11.68 & $-19.01,-19.41$ & 14.2 & 7.5 & 0.45\\
NGC~4516 &  VCC1479 & NGVSJ12:33:07.54+14:34:29.9 & SBa & 12.54,12.07 & $-18.54,-19.01$ & 23.6 & 5.7 & 0.15\\
IC~3167 & VCC407 & NGVSJ12:20:18.77+09:32:43.2 & SB0 & 14.00,13.48 & $-17.09,-17.61$ & 11.8 & 1.2 & $>2.0$ \\
\hline                
\end{tabular}
\end{center} 
\begin{tablenotes}
      \item \textbf{Notes.} (1) Galaxy name. (2) Alternative name from the Virgo Cluster Catalogue \citep[VCC,][]{Binggeli1985}. (3) NGVS reference name. (4)  Morphological type from the Extended Virgo Cluster Catalogue  \citep[EVCC,][]{kim2014}. (5) Apparent $r$-band magnitude from EVCC and based on Sloan Digital Sky Survey (SDSS\footnote{\url{https://www.sdss.org/}}) images and apparent $i$ band magnitude from NGVS \citep{Ferrarese2020}, respectively. (6) Total absolute $r$- and $i$-band magnitudes, assuming a distance of 16.5 Mpc, as done by \cite{Roediger2017} for all the galaxies included in the NGVS. (7) Half-light $i$-band radius from NGVS for NGC~4483 and NGC~4516 and from SDSS for IC~3167. (8) Stellar mass from NGVS, measured via modelling the galaxy spectral energy distribution (SED), following \cite{Roediger2017} with NGVS data. (9) Projected galaxy surface density, $\Sigma_5=5/(\pi d_5^2)$, where $d_5$ is the projected distance of the fifth nearest neighbor galaxy in Mpc. The neighbor galaxies were chosen from the SDSS to have a luminosity $M_r<-18$~mag and a radial velocity difference of less than 1000~km~s$^{-1}$ with respect to the reference galaxy, as done by \cite{Balogh2004}.
\end{tablenotes}
\end{table*}

\section{Broad-band images}
\label{sec:images}

\subsection{Images acquisition and reduction}

We use the $i$ and $g$ band images of the two galaxies taken from the NGVS, a deep and comprehensive optical imaging survey of the Virgo Cluster \citep{Ferrarese2012}, performed with the MegaCam instrument on the Canada–France–Hawaii Telescope (CFHT), reaching a surface brightness limit of $\mu_i\sim27.4$ ($\mu_g\sim29.0$)~mag~arcsec$^{-2}$, and with a pixel scale of $0.187$~arcsec~pixel$^{-1}$. The images are scaled to a photometric zero point of 30.0, such that AB magnitudes are given by $m({\rm AB})=−2.5\times \log({\rm DN}) + 30.0$, where DN is the number of counts measured in the frame. Further details can be found in \cite{Ferrarese2012}.

\subsection{Isophotal analysis}
\label{sec:iso_anal}

We perform the isophotal analysis of the sky-subtracted images of the sample using the \textsc{iraf} task \textsc{ellipse} on the $i$ band images. We fix the centre of the isophotes after checking that it does not vary within the uncertainties. The method measures the deviation of the isophotes from a perfect ellipse using a Fourier series.

We derive the radial profiles of the azimuthally-averaged surface brightness, $\mu_i$, ellipticity, $\epsilon$, position angle, PA, centre coordinates, and Fourier components up to the radius at which the error on the surface brightness as derived from the ellipse fitting is comparable to the $1\sigma$ error associated to the sky level. Then, we run the ellipse fitting on the $g$ band images adopting the same geometrical properties of the ellipses defined for the $i$ band (semi-major axis, $\epsilon$, PA, and centre coordinates), and build the $g-i$ colour profile. Features and changes along the colour profile may mark the transition between the regions dominated by different structural components \citep[i.e. the bulge, bar, and disc;][]{Peters2018}.

\begin{figure*}
    \centering
    \includegraphics[scale=0.75]{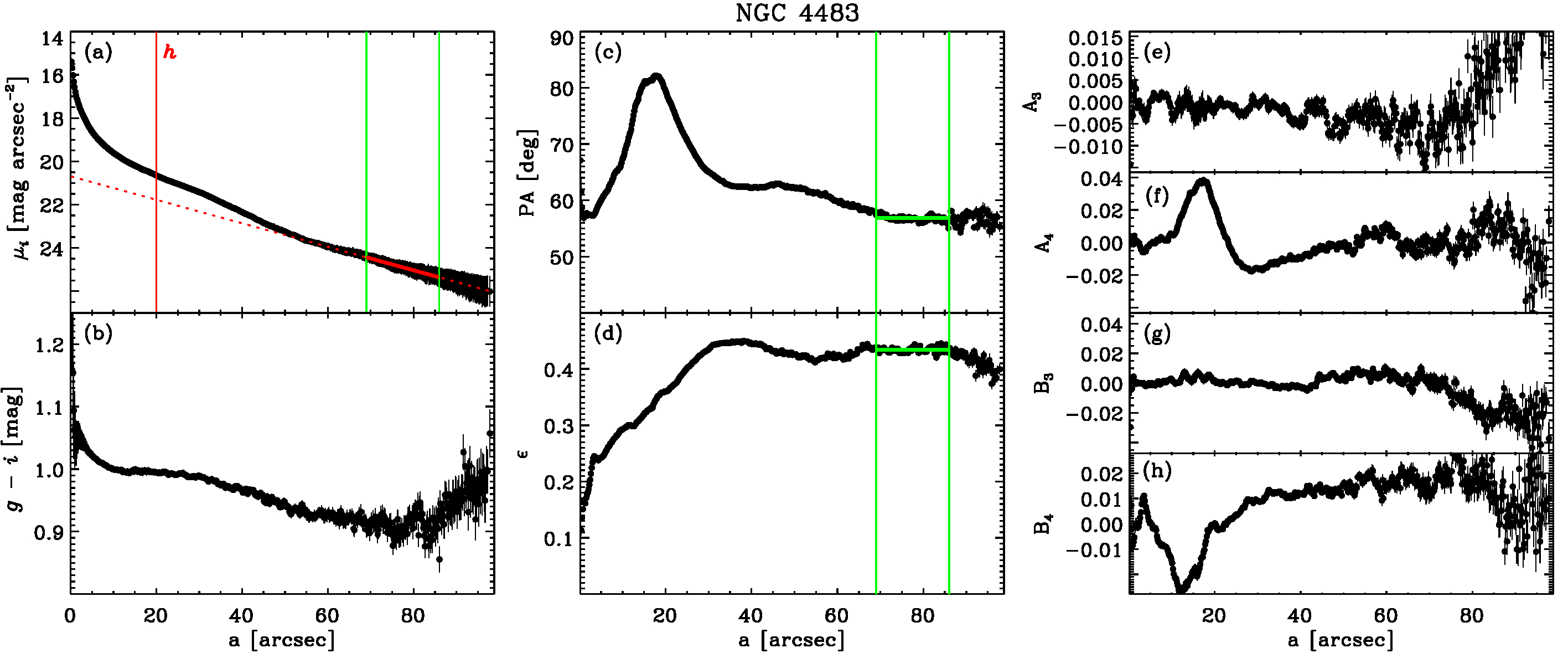}\\
    \hspace{+0.5cm}
    \includegraphics[scale=0.75]{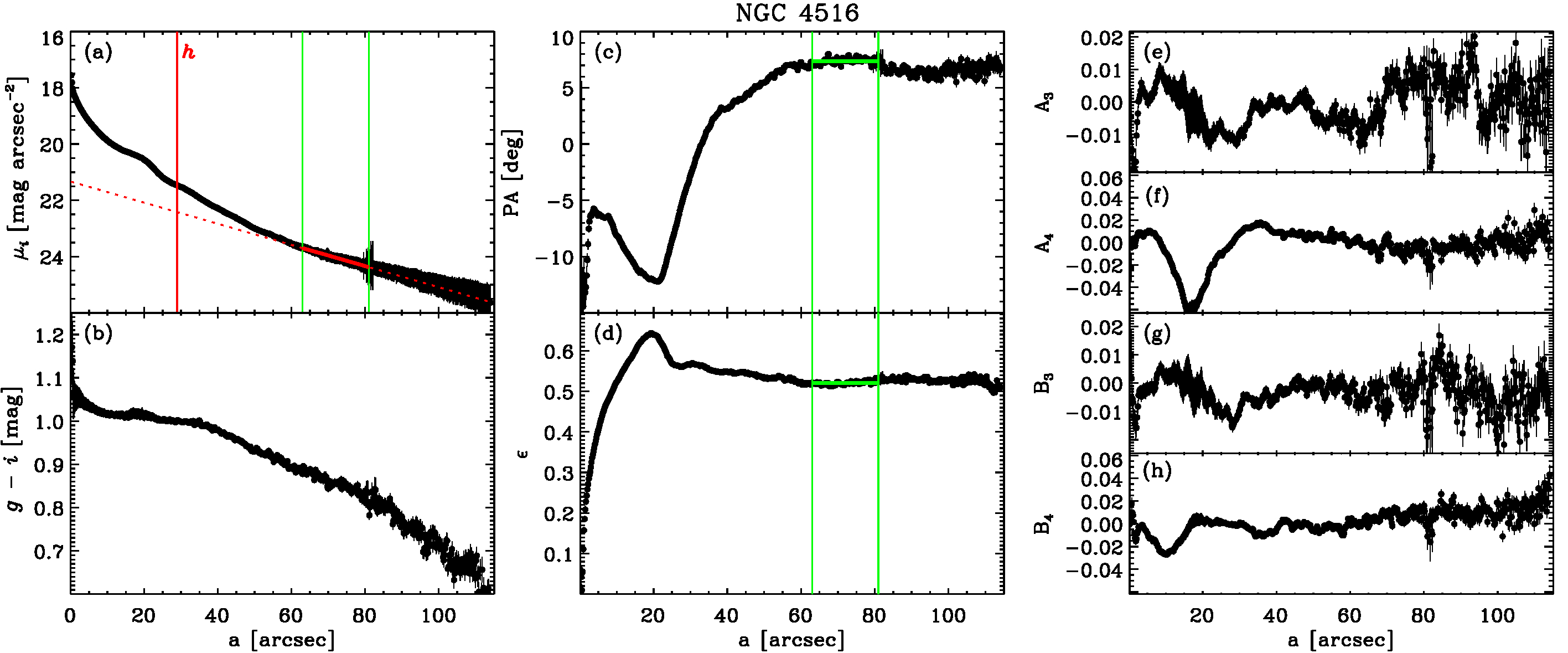}
    \caption{Isophotal analysis of the NGVS $g$- and $i$-band images of NGC~4483 (top panels) and NGC~4516 (bottom panels). For each galaxy we show the radial profiles of the surface brightness (a) and $g-i$ colour (b), PA (c) and $\epsilon$ (d), and deviations from ellipses $A_3$ (e), $A_4$ (f), $B_3$ (g) and $B_4$ (h). The vertical green lines (panels a,c,d) mark the region of the disc used to derive the disc parameters, while the red line (panel a) marks the best-fitting disc surface brightness with the solid portion corresponding to the region used to fit the disc exponential profile. The vertical red solid line marks the derived disc scale-length (panel a).}
    \label{fig:photometry_n4483}
\end{figure*}

Figure~\ref{fig:photometry_n4483} shows the radial profiles of $\mu_i$, $g-i$, $\epsilon$, PA, and deviations from ellipses $A_3$, $B_3$, $A_4$, and $B_4$ for the target galaxies. The photometric radial profiles derived here are in agreement with the results of the photometric analysis performed by \cite{Ferrarese2012,Ferrarese2020}.  

The radial profile of $\epsilon$ in an unbarred galaxy usually rises going from the central to the outer regions and flattens in the disc region. Barred galaxies usually show a similar behaviour at large radii, whereas in the inner regions the profile of $\epsilon$ presents a local peak and a sudden decrease in the region of the disc. In turn, the radial profile of PA in a barred galaxy is constant in the bar and disc regions, with generally two different values. The presence of a local maximum in the $\epsilon$ radial profile, typically associated with a nearly constant PA, is the signature of the bar. These behaviours are due to the shape and orientation of the bar-supporting orbits of the $x_1$ family \citep{Contopoulos1981,Wozniak1991,Aguerri2000}. 

The isophotal outer profiles allow to derive the geometrical properties of the galaxy disc. We derive the disc $\epsilon$ and PA for our target galaxies from the constant isophotal profiles in the disc region. By comparing the morphology of the galaxies from the NGVS $i$ band images with the isophotal profiles, we visually identify the disc-dominated region, which is characterised by a constant $\epsilon$ and PA, and possibly a change in the slope of the $g-i$ colour. We define the extension of the radial ranges by fitting the PA measurements with a straight line and considered all the radii where the line slope is consistent with being zero within the associated $1\sigma$ error, following \cite{Cuomo2019}. The disc $\epsilon$ is the mean value calculated adopting the same radial range identified for the disc PA. The $1\sigma$ errors are calculated for the disc PA and $\epsilon$. We derive the galaxy inclination as $i = \arccos (1 - \epsilon)$ assuming an infinitesimally thin disc. The resulting values are reported in Table~\ref{tab:isophotal}.


We calculate the $g-i$ colours of the bar and disc. In particular, the bar colour is given by the mean value  within the bar region, identified by projecting onto the sky plane the value of the mean bar length \rmean\ from Sec.~\ref{sec:barrands} and assuming as ${\rm PA_{\rm bar}}$ the value measured at the location of the $\epsilon$ peak produced by the bar, while the disc colour consists in the mean value within the disc region identified as described above. The resulting colours are reported in Table~\ref{tab:isophotal}. Discs are generally more metal-poor than the inner region hosting the bars, therefore they have lower $g-i$ colours (i.e. they are bluer, \citealt{Carter2009}) with respect to the bars.

\begin{table}
\addtolength{\tabcolsep}{-1pt}

\begin{center}
\caption{\label{tab:isophotal} Parameters from isophotal analysis.}
\begin{tabular}{ccccc}
\hline
Galaxy & disc PA & disc $i$ & $(g-i)_{\rm bar}$ & $(g-i)_{\rm disc}$ \\
 & [deg] & [deg] & [mag] & [mag] \\
(1) & (2) & (3) & (4) & (5) \\ \hline
NGC~4483 & $56.6\pm0.4$ & $55.5\pm0.3$ & $1.014\pm0.044$  & $0.914\pm0.020$ \\
NGC~4516 & $7.4\pm0.3$ & $61.4\pm0.2$ & $1.025\pm0.025$ & $0.845\pm0.022$ \\
\hline
\end{tabular}
\end{center}
\begin{tablenotes}
      \item \textbf{Notes.} (1) Galaxy name. (2) Disc position angle. (3) Disc inclination. (4) Bar colour. (5) Disc colour.
\end{tablenotes}
\end{table}

\subsection{Bar length and strength}
\label{sec:barrands}

We obtain the length of the bar semi-major axis, \rbar, by analysing the $i$ band images from the NGVS. We apply three independent methods to derive \rbar, given the difficulty in identifying the bar edges of the structure and to address the limitations and systematics of each methodology \citep{Combes1993,Rautiainen1999,Athanassoula2002,Cuomo2021}. 

First, we perform a Fourier analysis of the azimuthal luminosity profile of the deprojected NGVS $i$-band image, as in \cite{Ohta1990}, to get $R_{\rm bar/interbar}$. The deprojection is obtained by stretching the original image along the minor axis of the galaxy by a factor equal to $1/\cos{i}$, where $i$ is the disc inclination, and while conserving the flux. The values of the disc PA and $i$ are obtained as explained in Sec.~\ref{sec:iso_anal}. We derive the azimuthal profiles of the amplitude of the $m = 0, 1, 2, 3, 4, 5,$ and 6 Fourier components and the bar length is adopted to be the full width at half maximum (FWHM) of the luminosity contrasts between the bar, $I_{\rm bar}$, and interbar intensity, $I_{\rm ibar}$, as a function of radial distance, defined as $I_{\rm bar}= I_0 + I_2 + I_4 + I_6$ and $I_{\rm ibar} = I_0 - I_2 + I_4 - I_6$ \citep{Aguerri2000}. The upper and lower $1\sigma$ errors associated with $R_{\rm bar/interbar}$ are obtained repeating the Fourier analysis using only one half of the image.

The second method measures \rpha\ based on the profile of the constant phase angle of the $m = 2$ Fourier component, $\phi_2$ \citep{Debattista2002}. Following \cite{Buttitta2022}, the position where the $\phi_2$ changes by $\Delta{\rm \phi_2}=10\degr$ from the $\phi_2$ of the ellipse with the maximum bar/interbar intensity value provides \rpha. We obtain the radial profiles of the $\phi_2$ with the Fourier analysis described before. The upper and lower $1\sigma$ errors associated with \rpha\ are obtained identifying the radial region among with the values of $\phi_2$ are consistent with $10\degr$ within the errors obtained from the Fourier analysis.

The third approach measures $R_{\rm PA}$ based on the analysis of the PA of the deprojected isophotal ellipses \citep{Debattista2002}. Usually, the galaxy isophotes show a constant PA profile in the bar and disc regions, with two different values, related to the orientation of the bar and line of nodes, respectively. We adopt as $R_{\rm PA}$ the position where the PA changes by $\Delta{\rm PA}=5\degr$ from the PA of the ellipse with the maximum $\epsilon$ value \citep{Cuomo2019b}. The upper and lower $1\sigma$ errors associated with $R_{\rm PA}$ are obtained by identifying the radial region along which the values of $\Delta{\rm PA}$ are compatible with $5\degr$ within the errors from the ellipse fitting. 

\begin{figure*}
    \centering
    \includegraphics[scale=0.584]{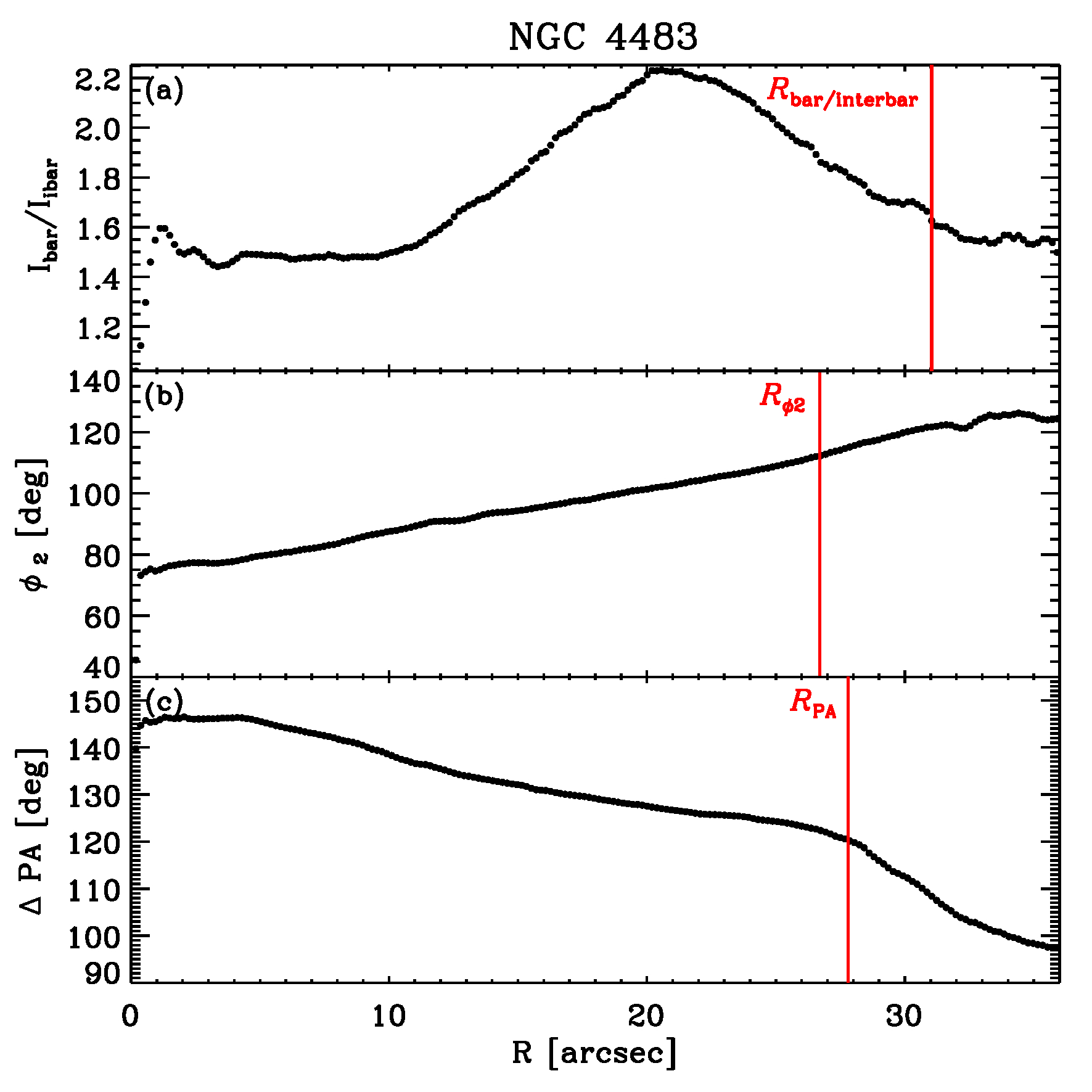}\hspace{+0.3cm}
    \includegraphics[scale=0.584]{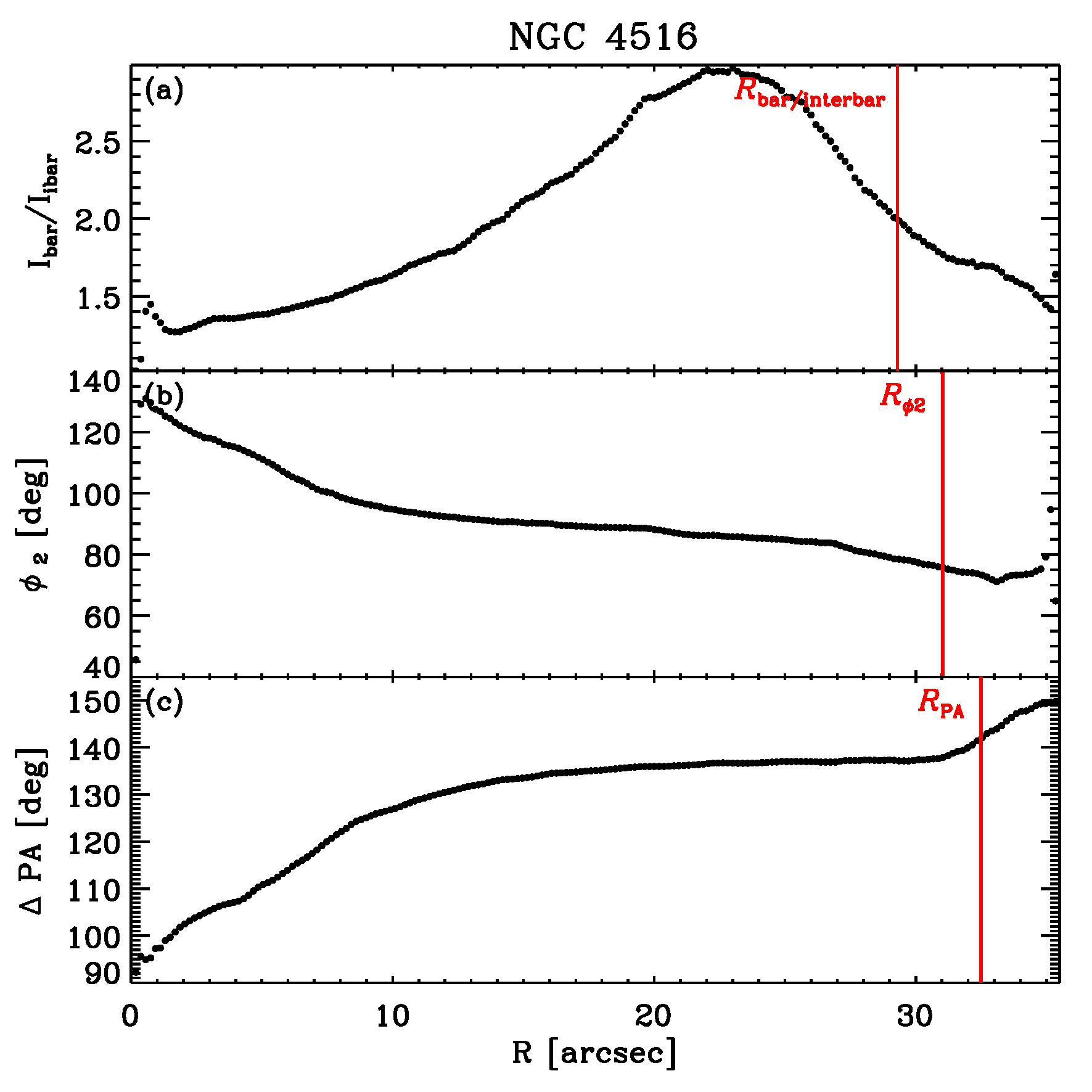}\hspace{-0.3cm}
    \caption{Bar length estimates from the NGVS $i$-band image of NGC~4483 (left panels) and NGC~4516 (right panels). We show the radial profiles of the bar/interbar intensity ratio (a), phase angle of $m=2$ Fourier component (b), and difference between the PA of the isophotes in the region of the $\epsilon$ peak and the bar (c). In each panel, the vertical red line marks the corresponding \rbar.}
    \label{fig:bar_radius}
\end{figure*}

Figure~\ref{fig:bar_radius} shows the three measurements of \rbar\ for the target galaxies. We then calculate the mean value, $R_{\rm bar}$, from the three measurements and adopt the lowest and highest deviations between $R_{\rm bar}$ and measurements as upper and lower $1\sigma$ errors on $R_{\rm bar}$, respectively. The bar region, i.e. the radial range between the shortest and largest bar length estimates from the different methods, is marked in panel (a) of Fig.~\ref{fig:gist} with two red ellipses. The measurements of \rmean, their mean value, and corresponding errors are reported in Table~\ref{tab:rbar}.

\begin{table*}
\begin{center}
\caption{\label{tab:rbar} Bar length and strength estimates.}
\begin{tabular}{ccccccc}
\hline
Galaxy  & $R_{\rm bar/interbar}$ & $R_{\phi_2}$ & $R_{\rm PA}$ & $R_{\rm bar}$ & $R_{\rm bar}$ & $S_{\rm bar}$ \\
 & {[}arcsec{]} & {[}arcsec{]}  &  {[}arcsec{]} & {[}arcsec{]} & [kpc] & \\
(1) & (2) & (3) & (4) & (5) & (6) & (7) \\ \hline
NGC4483 & $31.04^{+0.58}_{-0.01}$ & $26.7\pm0.4$ & $27.8\pm0.2$ & $28.5^{+2.5}_{-1.8}$ & $2.2^{+0.2}_{-0.1}$ & $0.378^{+0.001}_{-0.004}$ \\
NGC4516 & $29.3\pm0.3$ & $31.0^{+0.2}_{-0.4}$ & $32.5\pm0.2$ & $30.9\pm1.6$ & $2.9\pm0.1$ & $0.51^{+0.02}_{-0.01}$ \\


\hline
\end{tabular}
\end{center}
\begin{tablenotes}
      \item \textbf{Notes.} (1) Galaxy name. (2) Bar radius from Fourier analysis. (3) Bar radius from $\phi_2$ profile. (4) Bar radius from PA profile. (5) Mean bar length in arcsec. (6) Mean bar length in kpc. (7) Bar strength from Fourier analysis.
\end{tablenotes}
\end{table*}

We estimate the contribution of the bar to the gravitational potential of the galaxies in our sample, which we term the bar strength, \sbar, using the Fourier analysis. In particular, we adopt the peak of the ratio of the amplitudes of the $m = 2$ to $m = 0$ Fourier components, as in \cite{Athanassoula2002} and \cite{Guo2019}. The corresponding $1\sigma$ errors are obtained by repeating the Fourier analysis using the two half portions of the galaxy images, as in  \cite{Aguerri2015}. This parameter allows to distinguish between a strong (\sbar~$\ge 0.4$) and weak (\sbar~<0.4) bar \citep{Cuomo2019b}.

The values of \sbar\ are reported in Table~\ref{tab:rbar}.

\section{Integral-field spectroscopy}
\label{sec:muse}

\subsection{MUSE observations and data reduction}

The integral-field spectroscopic observations of NGC~4483 and NGC~4516 were carried out with the MUSE instrument of the ESO. MUSE was configured in wide field mode to ensure a FOV of $1\times1$~arcmin$^2$ with a spatial sampling of 0.2~arcsec~pixel$^{-1}$. The MUSE spectroscopic range covers 4800 to 9300~\AA\ with a spectral sampling of 1.25~\AA~pixel$^{-1}$ and an average nominal spectral resolution with a FWHM = 2.51~\AA\ \citep{bacon2010}. The observations were performed between January and April of 2021 and consisted of a central pointing ($\sim30$~min) on the galaxy centre and two off-set pointings along the galaxy major axis at a distance of 20 arcsec eastward ($\sim30$~min) and westward ($\sim30$~min) from the galaxy nucleus, divided into three observing blocks for each object. Each pointing was oriented along the disc PA, derived through a preliminary analysis of SDSS images performed during the preparation of the observational proposal. The observing nights were clear and the request of a FWHM seeing < 1.4 arcsec was mostly fulfilled. 

We perform the data reduction as detailed in \cite{Cuomo2019}, using the standard MUSE pipeline in the \textsc{exoreflex} environment \citep[version 2.8.4,][]{Weilbacher2020}. It includes bias and overscan subtraction, flat fielding, wavelength calibration, determination of the line spread function, sky subtraction, and flux calibration. The sky contribution is quantified using an on-sky dedicated exposure. Then, we determine the effective spectral resolution and its variation across the FOV, and produce the combined datacube of the galaxy. The resulting sky-subtracted datacube is characterized by a residual sky contamination, so we further clean it using the Zurich Atmospheric Purge algorithm \citep{soto2016}.

\subsection{Stellar kinematics and circular velocity}

We measure the stellar line-of-sight velocity distribution (LOSVD) of the two target galaxies from the sky-cleaned datacubes using the Galaxy IFU Spectroscopy Tool \citep[\textsc{gist,}][]{Bittner2019}, a modular pipeline based on the \textsc{pPXF} and \textsc{GandALF} routines \citep{cappellari2004,gandalf}. The \textsc{gist} pipeline makes use of the Voronoi tessellation of \cite{cappellari2003} for spatially binning the datacube spaxels to increase the signal-to-noise (S/N) ratio of each spectrum and ensure a reliable extraction of the relevant kinematic parameters. The S/N in each spaxel is obtained using the spectral range centred on the absorption-line calcium triplet (CaT) at $\lambda\lambda8498,8542,8662$~\AA. The instrumental velocity dispersion of $\sigma_{\rm instr}=35.3\pm1.2$~km~s$^{-1}$ \citep{Pizzella2018} is well suited for measuring the stellar kinematics in dwarf galaxies \citep[e.g.,][]{Hunter2005,Bidaran2020}. We adopt a target S/N ratio of 40 per pixel in order to be able to measure stellar velocity and velocity dispersion out to the spatial bins in the disc, while the spaxels mapping the central regions remained generally unbinned since their S/N largely surpassed the target value. We exclude from the fitting procedure the wavelength ranges with a spurious signal coming from imperfect subtraction of cosmic rays and residuals of the sky emission lines.

\begin{figure*}
    \centering   \includegraphics[scale=0.515]{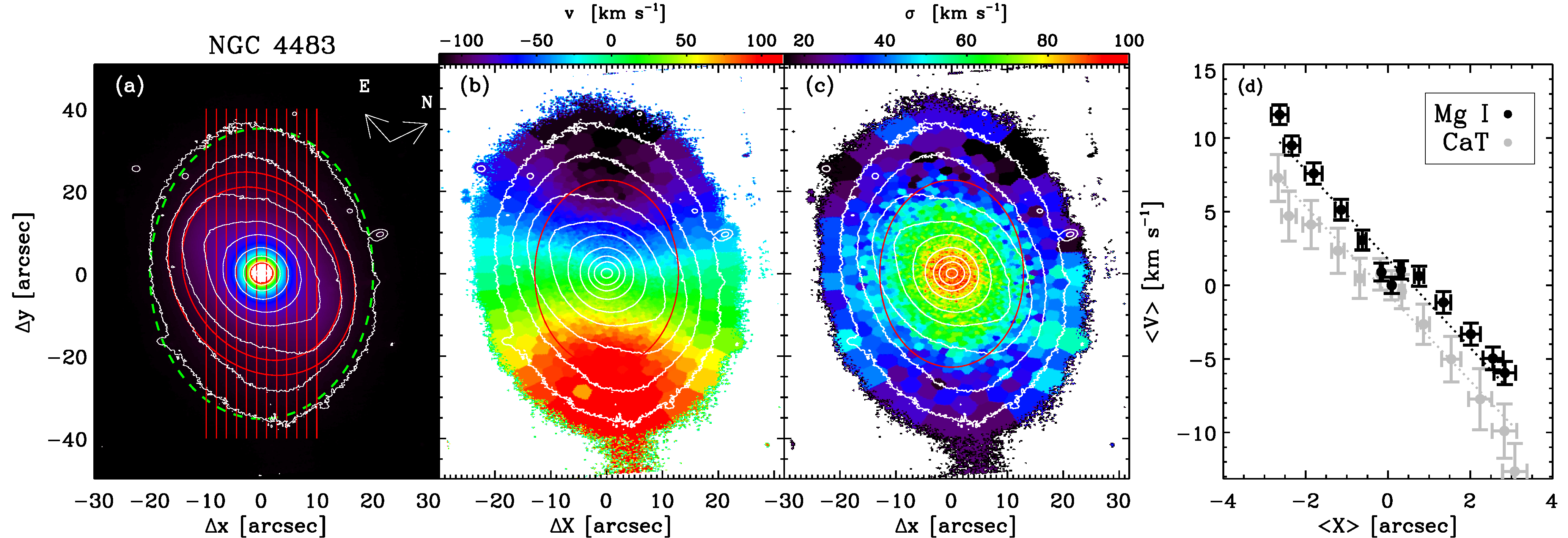}\vspace{0.5cm}  \includegraphics[scale=0.628]{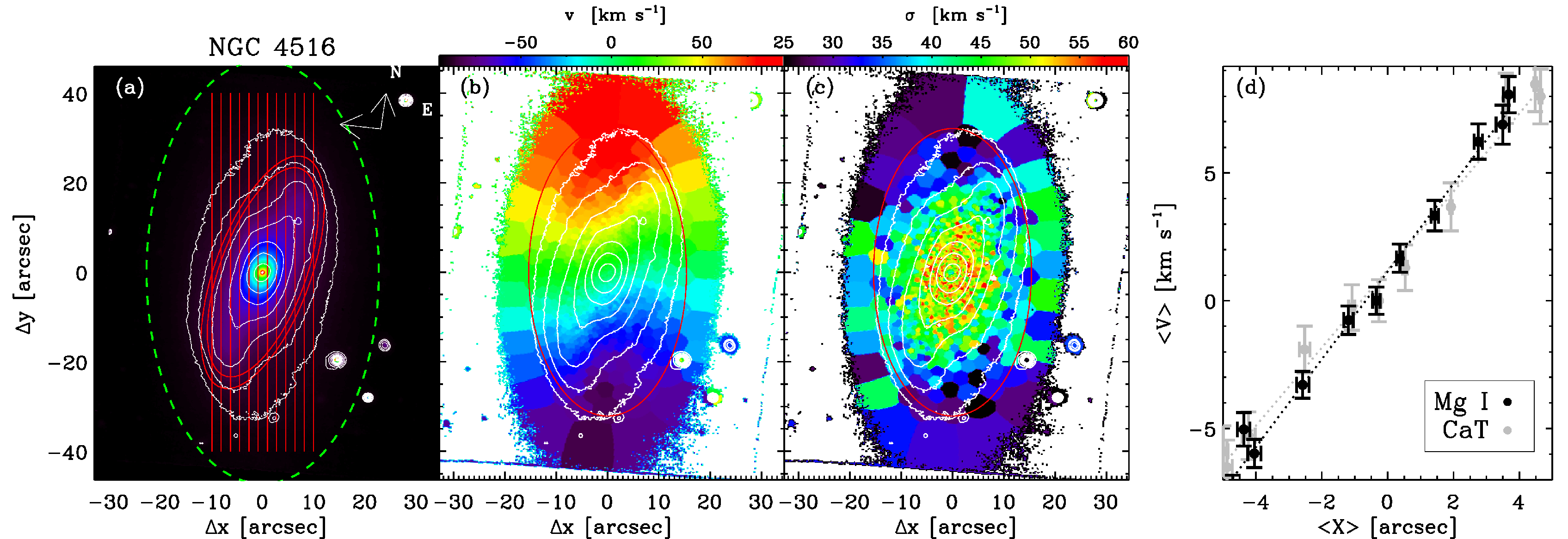}
    \caption{Stellar kinematics and bar pattern speed from MUSE data of NGC~4483 (top panels) and NGC~4516 (bottom panels). Panel (a): MUSE reconstructed image. The white lines mark a few isophotes to highlight the orientation of the bar and disc. The red ellipses mark the radial range between the shortest and longest bar length from the different methods used to derive \rbar. The green dashed ellipse marks the extension of the projected \rcor. The vertical red lines show the location of the pseudo-slits adopted to derive \omegabar. The disc major axis is parallel to the vertical axis. The North and East orientations are marked with white arrows. Panel (b): Map of the LOS stellar velocity subtracted of the systemic velocity. The value of \vcirc\ was derived using the spatial bins outside the red ellipse. Panel (c): Map of the LOS stellar velocity dispersion corrected for instrumental velocity dispersion. Panel (d): Kinematic integrals plotted as a function of photometric integrals measured using the spectral region of Mg I (black points) and CaT (grey points). The black/grey dotted lines represent the corresponding best fit to the data.}
    \label{fig:gist}
\end{figure*}

\begin{figure*}
    \centering   \includegraphics[scale=0.74]{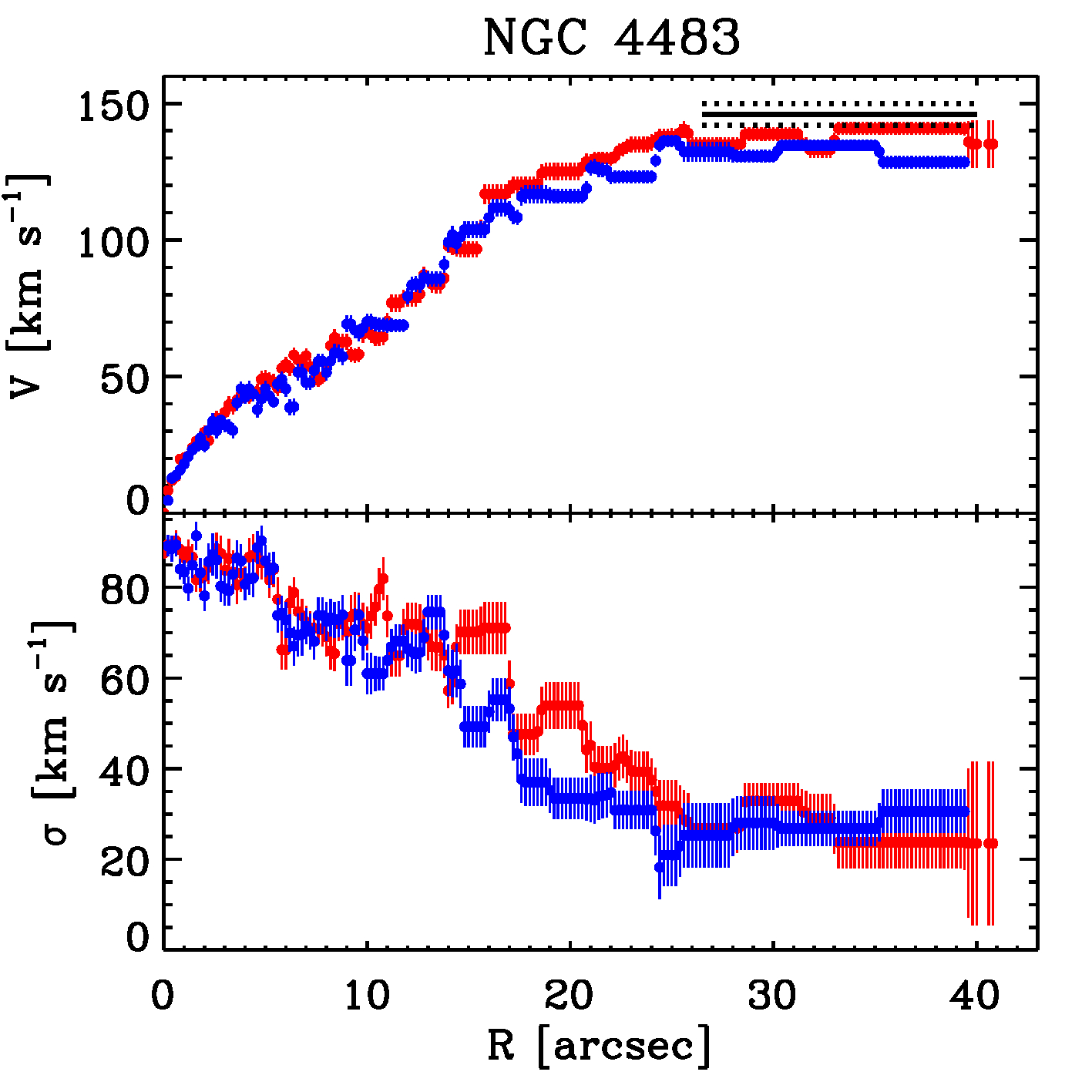}    \includegraphics[scale=0.74]{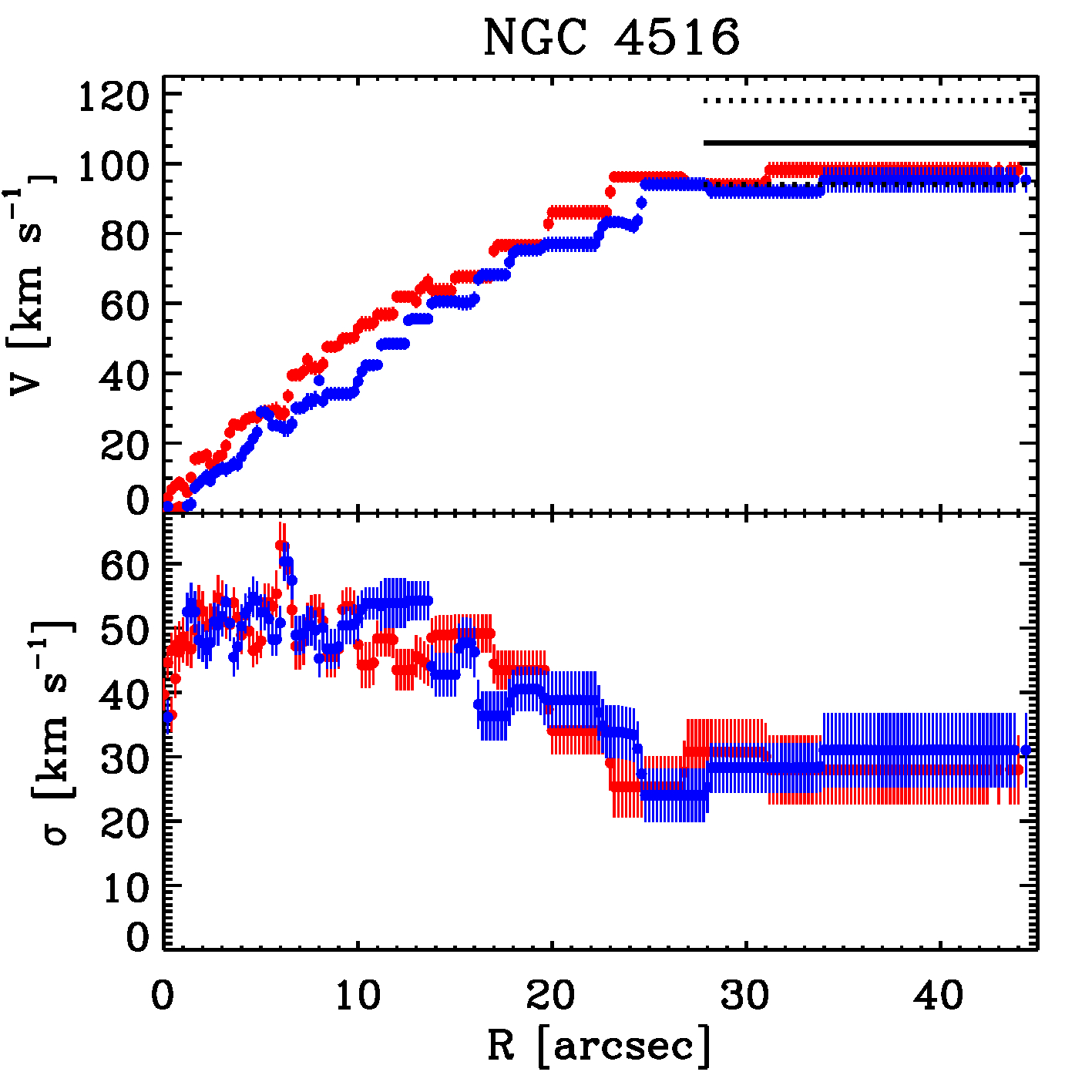} \\
    \caption{Stellar kinematics extracted along the disc major axis of NGC~4483 (left panels) and NGC~4516 (right panels) from the Voronoi-binned maps after being folded and deprojected onto the galaxy disc. Top panels show the stellar velocity deprojected on the disc plane. The horizontal black segments mark the circular velocity (solid line) and the $1\sigma$ error (dotted lines). The size of the segments mark the radial range used to derive \vcirc. The bottom panels show the stellar velocity dispersion. In each panel, the approaching and receding side of the disc are marked as red and blue points, respectively.}
    \label{fig:rot_curves}
\end{figure*}

To derive the stellar LOS velocity and velocity dispersion maps, we use the simple stellar population models based on the empirical X-shooter Spectral Library \citep[\textsc{xsl},][]{Verro2022}, characterised by a $\sigma_{\rm instr}\sim15$~km~s$^{-1}$ in the optical range. The errors on the kinematic parameters are estimated using Monte Carlo simulations and reach maximum values of $\sim5$~km s$^{-1}$ and $\sim15$~km s$^{-1}$ for the velocity and velocity dispersion, respectively. The best-fitting LOS velocity and velocity dispersion maps are shown in the central panels of Fig.~\ref{fig:gist}, while the stellar velocity and velocity dispersion profiles extracted along the disc major axis from the Voronoi-binned maps are shown in Fig.~\ref{fig:rot_curves}, after being folded and deprojected along the galaxy disc.

We derive the circular velocity \vcirc\ from the stellar LOS velocity and velocity dispersion in the disc region of our target galaxies using the asymmetric drift equation \citep{Binney2008}, as done by \cite{Cuomo2019}. We select the spatial bins outside the bar dominated region, which we identify by projecting \rmean\ to the sky plane. We derive \vcirc\ assuming the epicyclic approximation, where we adopt the ratio of the tangential-to-radial velocity dispersion components $\sigma_\theta/\sigma_R = 0.5$ in the galaxy plane for a constant circular velocity, as

\begin{equation}
    V_{\rm circ}^2 = (v-v_{\rm syst})^2 + \frac{\sigma^2}{\sin{i}^2(1+2\alpha^2\cot{i}^2)}\left[2R\left(\frac{1}{h}+\frac{2}{a}\right)-1\right]
\end{equation}

\noindent where $R$, $v_{\rm syst}$, $v$, and $\sigma$ are the distance with respect to the galaxy centre, galaxy systemic velocity, stellar LOS velocity, and LOS velocity dispersion projected along the disc major axis, assuming the disc PA and $i$ derived in Sec.~\ref{sec:iso_anal}, respectively. The disc scale-length, $h$, is obtained by fitting an exponential law to the surface brightness radial profile in the disc-dominated region for each galaxy, identified as explained in Sec.~\ref{sec:iso_anal}, while the corresponding $1\sigma$ error is the associated root mean square error. An exponential law for the stellar LOS velocity dispersion within the disc is assumed as well, with a scale-length $a$. Moreover, $\alpha=\sigma_z/\sigma_R$ represents the ratio between the perpendicular and radial disc velocity dispersion components, and is assumed to vary according to the galaxy morphological type \citep{Gerssen2012}. In particular, we use $\alpha = 0.85 \pm 0.15$ for NGC~4483 and $\alpha = 0.86 \pm 0.24$ for NGC~4516, adopting the EVCC morphological types (Table~\ref{tab:galaxies_poperties}). 

The $1\sigma$ error on \vcirc\ is calculated with a Monte Carlo simulation, allowing $i$, $h$, and $v_{\rm syst}$ to vary with a Gaussian distribution centred on the reference value and with a root mean square equal to the $1\sigma$ error of each parameter.

\begin{figure}
    \centering
    \includegraphics[scale=1.12]{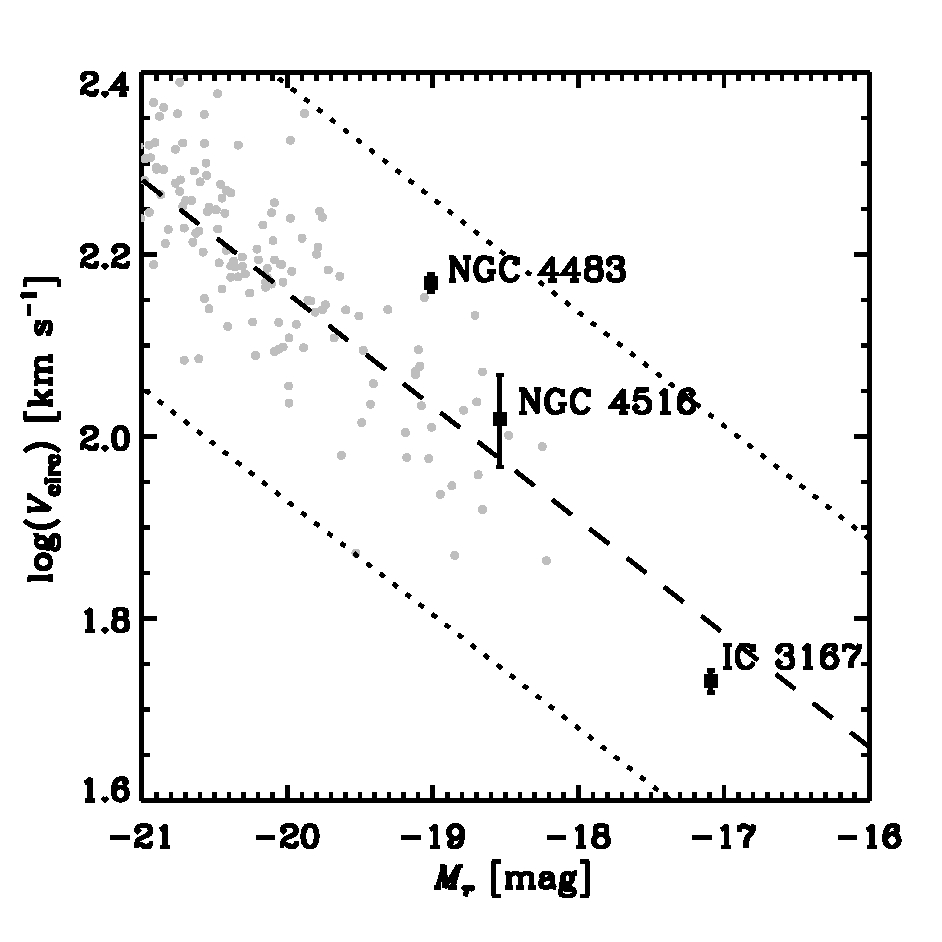}
    \caption{Tully-Fisher relation for our dwarf barred galaxies (black squares) and galaxy sample of Reyes et al. (2011) and Cuomo et al. (2019b) (grey points). The dashed line is the best-fitting relation from Reyes et al. (2011) and dotted lines bracket the region of $3\sigma$ deviation in $\log (V_{\rm circ})$.}
    \label{fig:tully_fisher}
\end{figure}

We check the reliability of our circular velocities by comparing the \vcirc\ values with the prediction of the Tully–Fisher relation. The \vcirc\ values of NGC~4483, NGC~4516, and IC~3167 are consistent within the $3\sigma$ scatter of the relation between the circular velocity and absolute SDSS $r$-band magnitude calculated by \cite{Reyes2011} for a sample of $\sim200$ nearby SDSS galaxies, which covers the range in magnitude of our dwarf galaxies (Fig.~\ref{fig:tully_fisher}).

The derived disc scale-length $h$ (Fig.~\ref{fig:photometry_n4483}), galaxy circular velocity \vcirc, and maximum value of the stellar velocity from the flat rotation curve projected on the galaxy plane, $V_{\rm flat}$, of our dwarf galaxies are reported in Table~\ref{tab:kin}. Low mass galaxies are expected to have low stellar velocity dispersion. For our dwarf galaxies, the stellar velocity dispersion in the disc region is between $\sim30$ and 40~km~s$^{-1}$. Therefore, the asymmetric drift correction is not large and the derived circular velocity is only $\sim10\%$ larger than the deprojected maximum stellar velocity $V_{\rm flat}$ (Fig.~\ref{fig:rot_curves}), in agreement with the results of \cite{Corsini2007} for NGC~4431.

\begin{table*}
\begin{center}
\caption{\label{tab:kin} Kinematic measurements of the sample of galaxies.}
\begin{tabular}{ccccccccc}
\hline
Galaxy  & $h$ & $V_{\rm flat}$ & $V_{\rm circ}$ & $\Omega_{\rm bar,Mg~I}$ & $R_{\rm cor}$ & $R_{\rm cor}$ & $\Omega_{\rm bar,CaT}$  & $\Omega_{\rm bar}$ \\
 & [arcsec] & [km s$^{-1}$] & [km s$^{-1}$] & [km s$^{-1}$ arcsec$^{-1}$] & [arcsec] & [kpc] & [km s$^{-1}$ arcsec$^{-1}$] & [km s$^{-1}$ kpc$^{-1}$] \\
(1) & (2) & (3) & (4) & (5) & (6) & (7) & (8) & (9)\\ \hline
NGC~4483 & $19.69\pm0.25$ & $140.9\pm3.0$ & $146\pm4$ & $3.6\pm0.2$ & $40.3\pm2.6$ & $3.2\pm0.2$ & $3.7\pm0.2$ & $45.2\pm2.8$ \\
NGC~4516 & $29.56\pm0.71$ & $98.2\pm6.0$ & $106\pm12$ & $2.0\pm0.1$ & $53.9\pm6.0$ & $4.3\pm0.5$ &  $2.0\pm0.1$ & $24.6\pm0.8$ \\
\hline
\end{tabular}
\end{center}
\begin{tablenotes}
      \item \textbf{Notes.} (1) Galaxy name. (2) Disc scale-length. (3) Maximum value of the flat stellar rotation curve. (4) Galaxy circular velocity derived from the asymmetric drift correction. (5) Bar pattern speed derived from the spectral region of the Mg I. (6) and (7) Bar corotation radius derived as $\Omega_{\rm bar,Mg~I}/V_{\rm circ}$ . (8) Bar pattern speed derived from the spectral region of the CaT. (9) Reference value of the bar pattern speed derived from the spectral region of the Mg I.
\end{tablenotes}
\end{table*}

\subsection{Bar pattern speed}

We derive $\Omega_{\rm bar}$ using the TW method on a tracer population of stars, which satisfies the continuity equation and assuming the bar is rotating as a rigid body. The simple idea is to observe the surface brightness and LOS velocity of the tracer along pseudo-slits located parallel to the disc major axis. When both position and velocity are measured with respect to the galaxy centre, then the luminosity-weighted mean velocity, $\langle V \rangle$,  divided by the luminosity-weighted mean position, $\langle X \rangle$, is equal to $\Omega_{\rm bar} \sin i$, where $i$ is the disc inclination. 

Following \cite{Cuomo2019}, the values of $\langle V \rangle$ are obtained collapsing for each pseudo-slit the datacube along the spatial directions and measuring the LOS velocity, $V_{\rm LOS}$, from each derived one-dimensional spectrum using \textsc{gist}, which is equivalent to measuring

\begin{equation}
    \langle V \rangle=\frac{\sum_{(x,y)} V_{\rm LOS} F(x,y)}{\sum_{(x,y)} F(x,y)}
\end{equation}

\noindent where $(x, y)$ are the single pixels in each pseudo-slit, and $F(x, y)$ is the corresponding flux measured in each pixel in the collapsed data. The luminosity weight required in the original TW formulation is calculated given the spaxels with higher signal give higher contribution in the collapsed spectrum and consequently in the $V_{\rm LOS}$ determination of each pseudo-slit during the fitting procedure. The $1\sigma$ errors on the kinematic integrals are estimated using a Monte Carlo simulation, as described by \cite{Bittner2019}. 

Each $\langle X \rangle$ requires measuring the mean position of the stars by collapsing the datacube along the spectral direction and weighting it by the luminosity in each pseudo-slit, as

\begin{equation}
    \langle X \rangle=\frac{\sum_{(x,y)} F(x,y) {\rm dist}{(x,y)}}{\sum_{(x,y)} F(x,y)}
\end{equation}

\noindent where ${\rm dist}(x, y)$ is the distance of each pixel with respect to the centre of the pseudo-slit.

For each galaxy, we define a suitable number of pseudo-slits crossing the bar. We adopted a pseudo-slit width of 9 pixels (1.8 arcsec) to avoid seeing smearing effects. 

To correctly derive \omegabar, in principle the pseudo-slits should have an infinitely extended length. However, \cite{Merrifield1995} showed that it is possible to adopt a finite reference length, when this is sufficient to map a symmetric contribution of the galaxy disc along the pseudo-slits. This corresponds to identifying the minimum pseudo-slit length when the integrals start to converge to a constant value \citep{Zou2019}. We test the convergence of the kinematic integrals as the radial region where their values are constant taking into account the corresponding errors, and define the pseudo-slit reference length. We then adopt the radial range along which the convergence was demonstrated to define the $1\sigma$ errors on the photometric integrals, adopting the root mean square of their values for each pseudo-slit, as done by \cite{Buttitta2022}.

The photometric and kinematic integrals are measured within the same wavelength range \cite[e.g.][]{Aguerri2015}. A wavelength window with no prominent emission lines should be used in order to trace the signal of the stars. The wavelength range from 4800 to 5600~\AA\ and centred on the absorption-line magnesium triplet (Mg I) at $\lambda5167, 5173, 5184$~\AA\ is first adopted, following \cite{Cuomo2019} and \cite{Buttitta2022}. The kinematic integrals are then derived using \textsc{gist} and the  \textsc{miles} stellar library \citep{Vazdekis2010}, characterised by a FWHM = 2.3~\AA. Using the \textsc{fitexy} routine in \textsc{idl}, we fit the integrals with a straight line with a slope $\Omega_{\rm bar} \sin i$ (Fig.~\ref{fig:gist}).

We then repeat the application of the TW method using the region of the CaT to measure both integrals and obtained a second, independent estimate of $\Omega_{\rm bar,CaT}$, reported in Table~\ref{tab:kin}. The two values of the bar pattern speed measured for each galaxy are consistent within $1\sigma$ errors. We conclude that the results are stable regardless of the selected wavelength range.

We adopt \omegabar\ measured in Mg I spectral range and the galaxy circular velocity \vcirc\ to derive \rcor. These quantities are listed in Table~\ref{tab:kin}. 

\subsection{Bar rotation rate}

We calculate ${\cal R}=R_{\rm cor}/R_{\rm bar}$ for NGG~4483 and NGC~4516, assuming the three different \rbar\ and the corresponding mean value $R_{\rm bar}$ obtained in Sec.~\ref{sec:barrands} and the galaxy circular velocity \vcirc. We provide four estimates of \rr, with the reference value assumed to be the one for $R_{\rm bar}$. Moreover, we adopt the maximum value of the flat stellar rotation projected on the galaxy plane to derive the lower limit of the bar rotation rate, ${\cal R}_{\rm flat}$. The $1\sigma$ errors associated to the different estimates of \rr\ were calculated performing a Monte Carlo simulation, taking into account the errors on disc $i$, each \rbar\ value, \omegabar, and \vcirc\ (or $V_{\rm flat}$). Moreover, we calculate the probability of the bar to be ultrafast/fast/slow, according to the reference value of \rr, performing a Monte Carlo simulation and assuming a uniform distribution for \rbar\ and a Gaussian distribution for \omegabar\ and \vcirc. All these results are reported in Table~\ref{tab:rr}. 

\begin{table*}
\begin{center}
\caption{\label{tab:rr} Bar rotation rates for the sample of galaxies.}
\begin{tabular}{ccccccc}
\hline
Galaxy  & ${\cal R}_{{\rm bar/interbar}}$ & ${\cal R}_{\phi_2}$ & ${\cal R}_{\rm PA}$ & ${\cal R}$ & ultrafast/fast/slow & ${\cal R}_{\rm flat}$ \\
(1) & (2) & (3) & (4) & (5) & (6) & (7) \\ \hline
NGC~4483 & $1.30^{+0.03}_{-0.04}$ & $1.51^{+0.04}_{-0.04}$ & $1.45^{+0.09}_{-0.04}$ & $1.41^{+0.10}_{-0.11}$ & 0/52/48 & $1.37^{+0.10}_{-0.05}$ \\
NGC~4516 & $1.84^{+0.09}_{-0.09}$ & $1.74^{+0.09}_{-0.09}$ & $1.66^{+0.09}_{-0.09}$ & $1.75^{+0.10}_{-0.10}$ & 0/6/94 &  $1.62^{+0.10}_{-0.10}$ \\


\hline      
\end{tabular}
\end{center}
\begin{tablenotes}
      \item \textbf{Notes.} (1) Galaxy name. (2) Bar rotation rate assuming bar length $R_{\rm bar/interbar}$. (3) Bar rotation rate assuming bar length $R_{\phi_2}$. (4) Bar length $R_{\rm PA}$. (5) Bar rotation rate assuming the mean bar length $R_{\rm bar}$. (6) Per cent probability of having an ultrafast/fast/slow bar. (7) Bar rotation rate assuming the mean bar length $R_{\rm bar}$ and maximum stellar rotation velocity $V_{\rm flat}$.
\end{tablenotes}
\end{table*}

\section{Results}
\label{sec:results}

In this section, we present and discuss the results for NGC~4483 and NGC~4516. 

\subsection{NGC~4483}


NGC~4483 hosts a short ($R_{\rm bar}=2.1\pm0.3$~kpc) and weak bar ($S_{\rm bar}=0.378^{+0.001}_{-0.004}$), according to the classification criteria of \cite{Cuomo2019b}. The bar is associated with a clear peak in the PA profile, and a boxy shape of the isophotes (peaking at $B_4\sim-0.03$). The bar is embedded in an intermediate-inclination disc characterised by a constant profile of PA and $\epsilon$, and discy isophotes ($B_4\sim0.01$). The $(\mu_g-\mu_i)$ colour in the bar and disc regions is blue and increasing outwards, while it stars decreasing at $R > 80$~arcsec (Fig.~\ref{fig:photometry_n4483}). 

The stellar kinematics shown in Figs.~\ref{fig:gist} and \ref{fig:rot_curves} reveals a maximum of the LOS stellar velocity measured at $v\sim115$~km~s$^{-1}$. The LOS stellar velocity dispersion shows a central value of $\sim100$~km~s$^{-1}$, decreasing to $\sim30$~km~s$^{-1}$ in the disc outside the bar region.

We apply the TW method after defining 13 pseudo-slits crossing the bar, with half length of 40 arcsec. The bar has $\Omega_{\rm bar}=45.2\pm2.8$~km~s$^{-1}$~kpc$^{-1}$. This translates to ${\cal R}=1.41^{+0.10}_{-0.11}$, when considering \rmean. The nominal value of the bar rotation rate belongs to the slow regime for all the different bar length estimates, while the bar has a probability of being slow (fast) of 48\% (52\%). When we adopt $V_{\rm flat}$ to derive the lower limit for the bar rotation rate, we get ${\cal R}_{\rm flat}=1.37^{+0.10}_{-0.05}$. 

The $\langle X \rangle$ and $\langle V \rangle$ integrals are nicely aligned, except for the three central slits (Fig.~\ref{fig:gist}) for both the adopted wavelength ranges. This may be caused by the presence of a central extra component, rotating at a different pattern speed with respect to the main bar \citep{Corsini2003,Meidt2009}. The visual inspection and photometric analysis suggests this galaxy hosts a barlens and/or nuclear stellar disc, which also causes the non-zero value of the bar-interbar profile in the inner region of the galaxy (Fig.~\ref{fig:bar_radius}). 

\subsection{NGC~4516}

The isophotal analysis (Fig.~\ref{fig:photometry_n4483}) shows a peak in the $\epsilon$ radial profile, produced by the short ($R_{\rm bar}=2.3\pm0.3$~kpc) but strong bar ($S_{\rm bar}=0.51^{+0.02}_{-0.01}$). The inner isophotes are boxy peaking at $B_4\sim-0.03$ ($R<20$~arcsec), given the presence of the BP bulge. The disc has an exponential surface brightness profile ($R\sim60-80$~arcsec) with constant PA and $\epsilon$ values, and an increasing blue $g-i$ colour. In the outer region of the galaxy ($R>80$~arcsec), the $g-i$ colour steepens, whereas the PA and $\epsilon$ remain constant.

The LOS stellar velocity reaches a maximum constant value of $v=85$~km~s$^{-1}$. The LOS stellar velocity dispersion rises from $\sim30$~km~s$^{-1}$ in the disc to 60~km~s$^{-1}$ in the bar region. A weak $\sigma$-drop is observed in the very central region \citep[see also][]{Mendezabreu2014,Portaluri2017}, where the velocity dispersion drops to $\sim45$~km~s$^{-1}$ (Fig.~\ref{fig:gist}).

The measurement of \omegabar\ is obtained defining 11 pseudo-slits with half length of 40 arcsec. We find $\Omega_{\rm bar}=24.6\pm0.8$~km~s$^{-1}$~kpc$^{-1}$, which leads to ${\cal R}=1.75^{+0.10}_{-0.10}$, obtained adopting \rmean. Moreover, the bar is slow regardless of which of the four derived bar length estimates is used. The lower limit ${\cal R}_{\rm flat}=1.62^{+0.10}_{-0.10}$ is likewise in the slow regime. The bar in NGC~4516 has a 94\% (6\%) probability of being slow (fast).

Faint spiral arms are associated with the bar in NGC~4516. These spiral arms are therefore located well within the bar corotation radius (see Fig.~\ref{fig:gist}, bottom row, left panel), which can have some implications on the orbits supporting them \citep{Patsis2017}, but their study is beyond the scope of this paper.

\subsection{Relations among bar properties}

We explore possible relations among barred galaxies with a direct measurement of \omegabar\ based on the TW method. In particular, we compare the results from \cite{Cuomo2020,Cuomo2022}, \cite{Buttitta2022}, and \cite{Garma-Oehmichen2022}. In the following, we discuss the bar properties of IC~3167 together with those of NGC~4483 and NGC~4516. The barred galaxy IC~3167 originally presented by \cite{Cuomo2022} was observed within the same observational campaign of NGC~4483 and NGC~4516. It is a lenticular galaxy with stellar mass of $M_{\ast}=1.2\times 10^9~\msun$ \citep{Roediger2017}. The bar of this galaxy is quite peculiar, since it has a remarkable lopsided shape, as shown by the third sine Fourier coefficient peaking at $B_3\sim −0.05$ for the galaxy isophotes, and by the large values of the odd components of the Fourier analysis. Using MUSE IFS and SDSS $i$-band image, \cite{Cuomo2022} were able to characterise the bar properties and apply the TW method. The bar in IC~3167 is slowly-rotating, with ${\cal R}=1.7^{+0.5}_{-0.3}$. Given the peculiar nature of this bar and that IC~3167 is part of a group of galaxies in-falling in the Virgo Cluster, the formation of the lopsided bar could be induced by the ongoing gravitational interaction with the cluster or by an interaction within the group of in-falling galaxies it belongs to. 

We explore the relations previously studied by \cite{Cuomo2020} between the main parameters of the bars (\rbar, \sbar, \omegabar, \rcor, and \rr) and their host galaxies (Hubble type, $M_r$, and Petrosian radius \rpetro). In particular, we consider galaxies for which the TW method resulted in a \omegabar\ estimate with relative error smaller than 50\%, and we compare previous results to those for our dwarf galaxy sample.

\begin{figure*}
    \centering
    \includegraphics[scale=0.7]{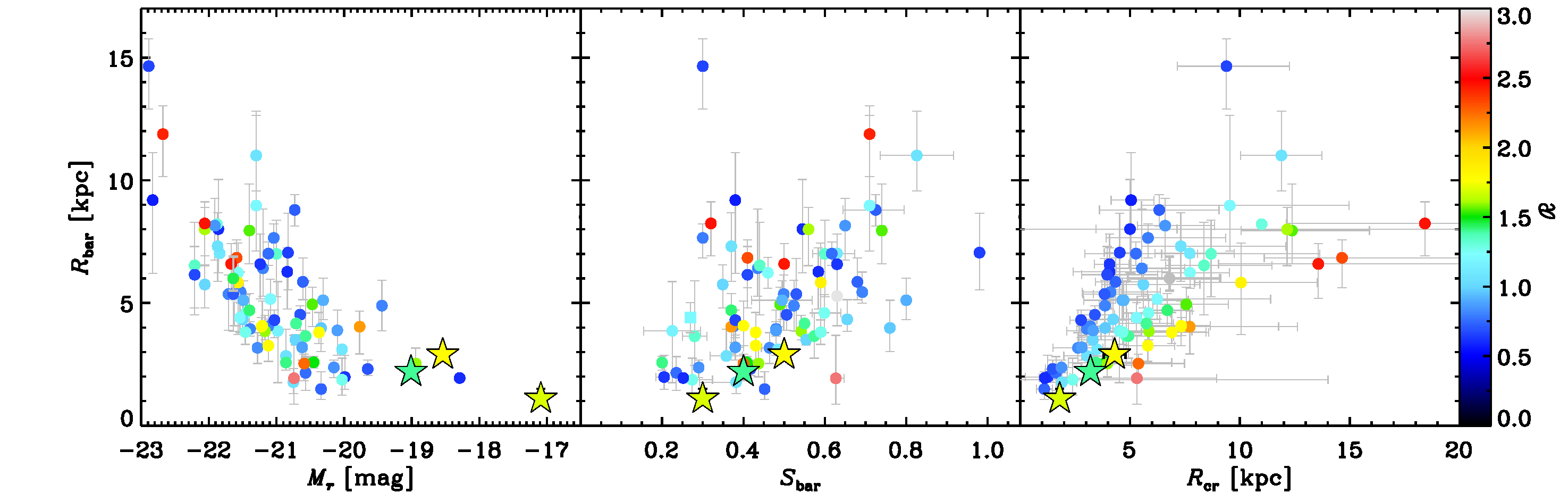}
    \includegraphics[scale=0.75]{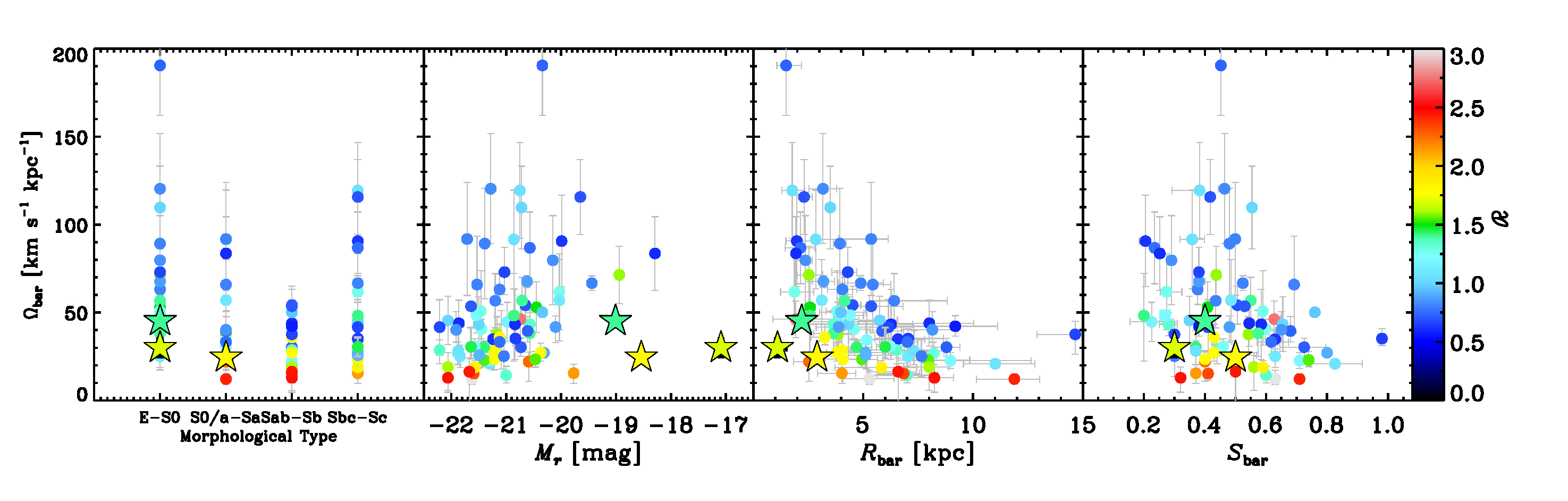}
    \caption{Relations among bar and galaxy properties of our dwarf galaxies (stars) and the sample of galaxies from Cuomo et al. (2020) (circles), with \omegabar\ derived using the TW method. The points are colour-coded according to their value of \rr.}
    \label{fig:relations}
\end{figure*}

Figure~\ref{fig:relations} shows the relations among bar and galaxy properties from the literature, with \omegabar\ derived using the TW method, after excluding bars with rotation rates consistent with the ultrafast regime. Dwarf galaxies have shorter and weaker bars, despite being early-type disc galaxies, which generally host long and strong bars \citep{Erwin2005}. Moreover, the bars in dwarf galaxies rotate with lower \omegabar\ and have \rr\ mostly in the slow regime. Similar correlation to those observed for the massive counterparts are observed when the bar lengths are normalised by \rpetro. 

\begin{figure}
    \centering
    \includegraphics[scale=0.75]{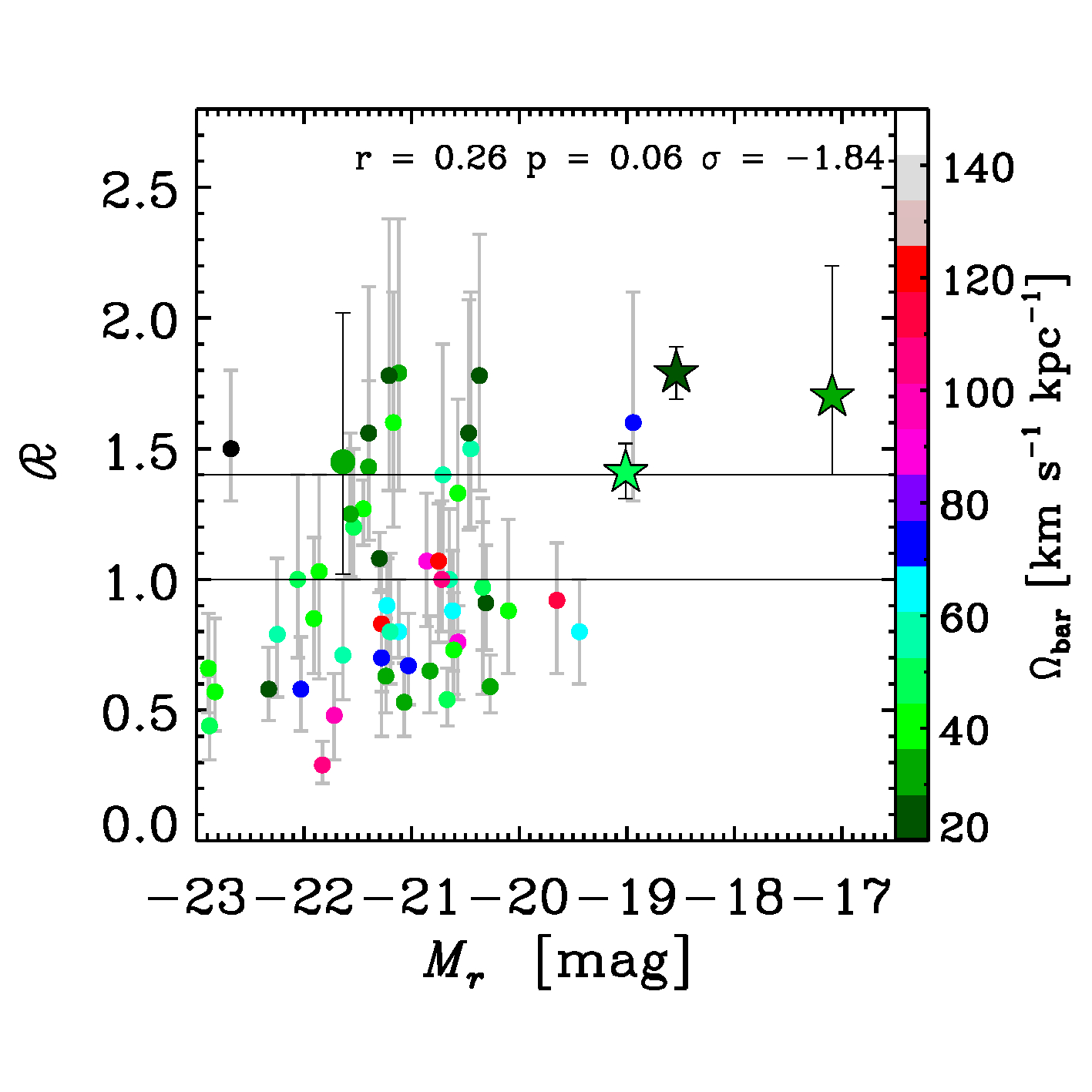}
    \caption{Bar rotation rates versus total absolute magnitude of the barred galaxies with $\Delta {\cal R}/{\cal R}<0.5$. The points are colour-coded according to the value of the bar pattern speed. The stars mark our dwarf barred galaxies. The small circles refer to results from the literature (Cuomo et al. 2020, Buttitta et al. 2022, and reference therein). The big circle represents the mean value for the sample of MaNGA MW analogues studied by Garma-Oehmichen et al. (2022).} 
    \label{fig:r_mr}
\end{figure}

We then explore the relation between \rr\ and galaxy stellar mass. We adopt $M_r$ as a proxy of $M_{\ast}$, since we do not have mass estimates for all the galaxies in the literature (Fig.~\ref{fig:r_mr}). Most of the analysed galaxies from the literature are massive. Bars in dwarf galaxies tend to rotate slower with respect to the other galaxies. When selecting galaxies with $\Delta {\cal R}/{\cal R}<0.5$, a weak correlation appears (with a Spearman rank correlation $r>0.2$ and the corresponding two-sided significance of its deviation from the null hypothesis $p <0.1$). However, the low mass regime is still too poorly explored to derive stronger conclusions.

\begin{figure}
    \centering
    \includegraphics[scale=0.78]{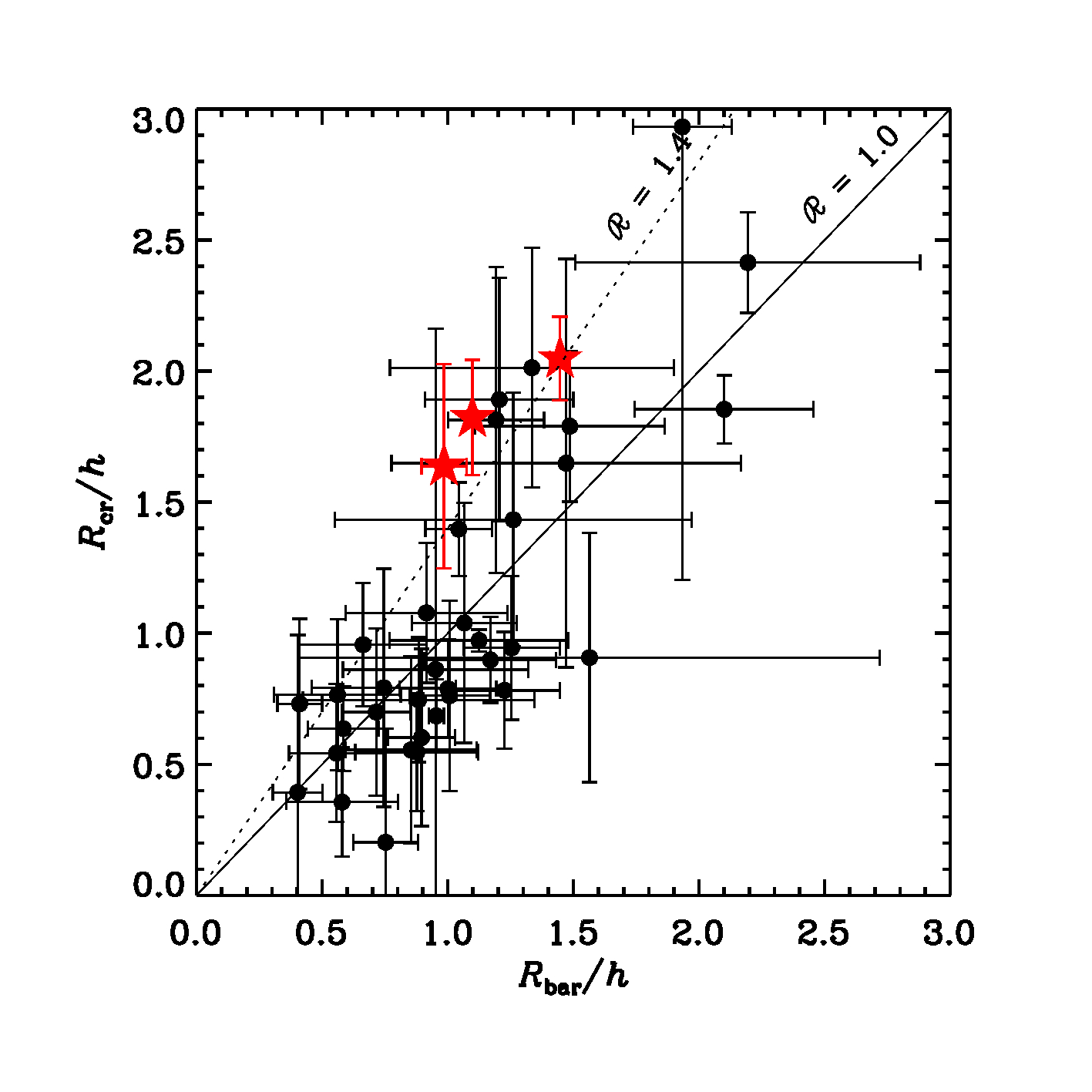}
    \caption{Ratio between the corotation radius and disc scale-length as a function of the ratio between the bar radius and disc scale-length for the dwarf barred galaxies (stars) and massive barred galaxies (circles) from literature \citep{Cuomo2019b}. The solid and dashed lines mark ${\cal R} = 1.0$ and $1.4$, respectively.}
    \label{fig:disc_param}
\end{figure}

To investigate the disc regions of our barred galaxies, we measure the ratios between \rcor\ and $h$ and between \rbar\ and $h$ to be compared to their massive counterparts from the literature (Fig.~\ref{fig:disc_param}). The bars in dwarf galaxies are slightly larger than their disc scale-lengths ($1.0<{R_{\rm bar}}/h<1.5$), whereas the corotation radii is located further out (${R_{\rm cor}}/h>1.5$). This result is different when compared to massive galaxies, where most of the bars and corotation radii are confined within or are close to their disc scale-length ($1.0<{R_{\rm bar}}, {R_{\rm cor}}/h<1.5$), with some exceptions in the weak bar regime (see \citealt{Cuomo2019b}, for a discussion).

\section{Discussion}
\label{sec:discussion}

\subsection{Issues related to bar length estimates}

The correct identification of the extension of the bar is crucial to derive the bar rotation rate and to compare observational results to the predictions from simulations. As an example, the accurate identification of \rbar\ was able to solve the problem of ultrafast bars, which were observed through TW analyses, but which are theoretically not permitted according to the stability of the stellar orbits supporting the bar. Indeed, \cite{Cuomo2021} showed ultrafast bars become actually fast when the bar length measurements used to derive \rr\ are measured carefully.

However, the edges of bars are not always sharp and easy to identify, given that these structures are often associated with other components (such as rings, pseudo-rings and/or spiral arms). Moreover, spiral arms are responsible for a large fluctuation of \rbar\ on a dynamical timescale depending on the strength of the spiral structure and on the method used to measure it \citep{Hilmi2020}. 

Various methods have been developed to derive \rbar. Each of them is based on a photometric analysis of a galaxy image and suffers from some limitations \citep[see][for a discussion]{Corsini2011}. In Sec.~\ref{sec:barrands} we use three different methods to derive \rbar. A Fourier decomposition of the galaxy light, which has been largely tested with $N$-body simulations and generally provide an error of 4 per cent on \rbar\ \citep{Athanassoula2002}. However, this method can be hampered by the presence of a non-axisymmetric disc and strong spiral arms \citep{Aguerri2000}. Then, we analyse the profile of the $\phi_2$ and PA of the deprojected isophotal ellipses, which are quite constant within the bar region. However, the presence of a prominent bulge may complicate the application of these methods \citep{aguerri2005}. 


Other possible methods can be adopted as well \citep[e.g.][]{mendezabreu2012,Lee2020}, each of them with its own issues. As an example, the radial profile of $\epsilon$ usually presents a local maximum in the bar region, which traces the shape and size of the stellar orbits supporting the bar, which has been used to derive the extension of the bar \citep{Wozniak1991}. However, this method underestimates the real \rbar\ because the peak in $\epsilon$ is typically located in a region within the bar end \citep{Erwin2017}. This is the reason why we have decided not to use it, despite having being used before \citep{Aguerri2015,Cuomo2019b}.


Given the difficulty in measuring \rbar\ for each galaxy analysed with the TW technique, here we calculated the bar rotation rates corresponding to each measurements of \rbar, and then the value corresponding to the mean bar length, $R_{\rm bar}$, as usually done in the literature. The derived rotation rates of the bars in NGC~4483 and NGC~4516 turn out to be larger than 1.4 regardless of the method used to derive \rbar, except for the shorter \rbar\ measurement of NGC~4483. These results translate into a probability of being fast/slow of 52/48\% for NGC~4483 and 6/94\% for NGC~4516 (Table~\ref{tab:rr}).

\subsection{Bar pattern speed using different wavelength ranges}

Deriving the bar rotation rate using different wavelength ranges has never been extensively tested. The application of the TW method could give different \omegabar\ when changing the wavelength range, as is the case for \rbar\ and \sbar. For these reasons, we have repeated the measurements of \omegabar\ using the CaT region centred at $\sim8000$~\AA\ and compared the results with the reference analysis obtained with the Mg I region. We are able to do this test after performing a careful sky subtraction in the MUSE datacubes. When residuals from the sky lines are still visible near the CaT lines, we properly mask them before measuring the $\langle X \rangle$ and $\langle V \rangle$ integrals. 

Our measurements of \omegabar\ from the spectral region around Mg I and CaT are shown in the right panels of Fig.~\ref{fig:gist}. For each galaxy, the derived \omegabar\ from the two different wavelength ranges are consistent within $1\sigma$ errors. This suggests that the correct bar pattern speed can be derived using the TW analysis regardless of the adopted wavelength range, after excluding gas and sky emission lines. This result may open the application of the TW method to measurements from different spectrographs in different wavelength regimes.

\subsection{Bar properties in dwarf galaxies}


Bars have been generally thought to be rare in dwarf galaxies \citep{MendezAbreu2010,mendezabreu2012}, but recent techniques have been developed to highlight that they can host disc structures including bars \citep{Michea2021}. The TW method has been applied to a few dwarf galaxies so far \citep{Corsini2007,Cuomo2022}. This low number makes it impossible to obtain strong conclusions on the formation of bars in dwarf galaxies.

Our work is a first step towards a more complete characterisation of bars in dwarf galaxies, showing that the high quality of MUSE data together with a careful photometric analysis allows to push the application of the TW method to the low-mass galaxy regime. 


The objects analysed here are lenticular or early-type spiral galaxies and generally host short ($R_{\rm bar}=1-2$~kpc) and intermediate-to-weak ($S_{\rm bar}=0.3-0.5$) bars. Their bar pattern speeds are low ($\Omega_{\rm bar}\sim30-45$~km~s$^{-1}$~kpc$^{-1}$). The bar rotation rate is $>1.4$ for IC~3167, NGC~4483, and NGC~4516, with a probability of being slow always larger than $\sim70$\% for IC~3167 and NGC~4516. NGC~4483 has an equal probability of being either fast or slow.


Bar pattern speed and rotation rates in dwarf galaxies tend to be slower than in massive counterparts. The expected secular evolution of bars within the $\Lambda$CDM cosmology suggests these structures should evolve by exchanging angular momentum with the other components while slowing down due to the dynamical friction exerted by the DM halo \citep{Debattista2000,Athanassoula2013}. Both these phenomena should slow down the bar pattern speed, pushing the bar rotation rate towards the slow regime, while the bar grows in length and strength. This seems to be the case for the analysed dwarf barred galaxies. 

Dwarf galaxies are expected to host centrally-concentrated DM halos \citep{Cote1991}, which should slow down the bar rotation through dynamical friction \citep{Debattista2000}. Two out of three dwarf galaxies analysed here host genuine slowly rotating bars, in agreement with the prediction of $\Lambda$CDM cosmology. However, to robustly confirm this scenario, it would be necessary to derive the DM content within the bar region as recently done by \cite{Buttitta2023} for two massive barred galaxies. 

Finally, both NGC~4483 and NGC~4516 show evidence of a BP bulge in the inner part of their bars. BP bulges are expected to be rare in such low-mass galaxies \citep{Erwin2017} and to be associated to a slowly rotating boxy bar \citep{Chaves-Velasquez2017}. Studying the properties of these structures will be useful to characterise the orbit supporting them, and to understand the mechanisms involved in their formation.

\subsubsection{Environmental effects}

\begin{figure}
    \centering
    \includegraphics[angle=90,scale=0.385]{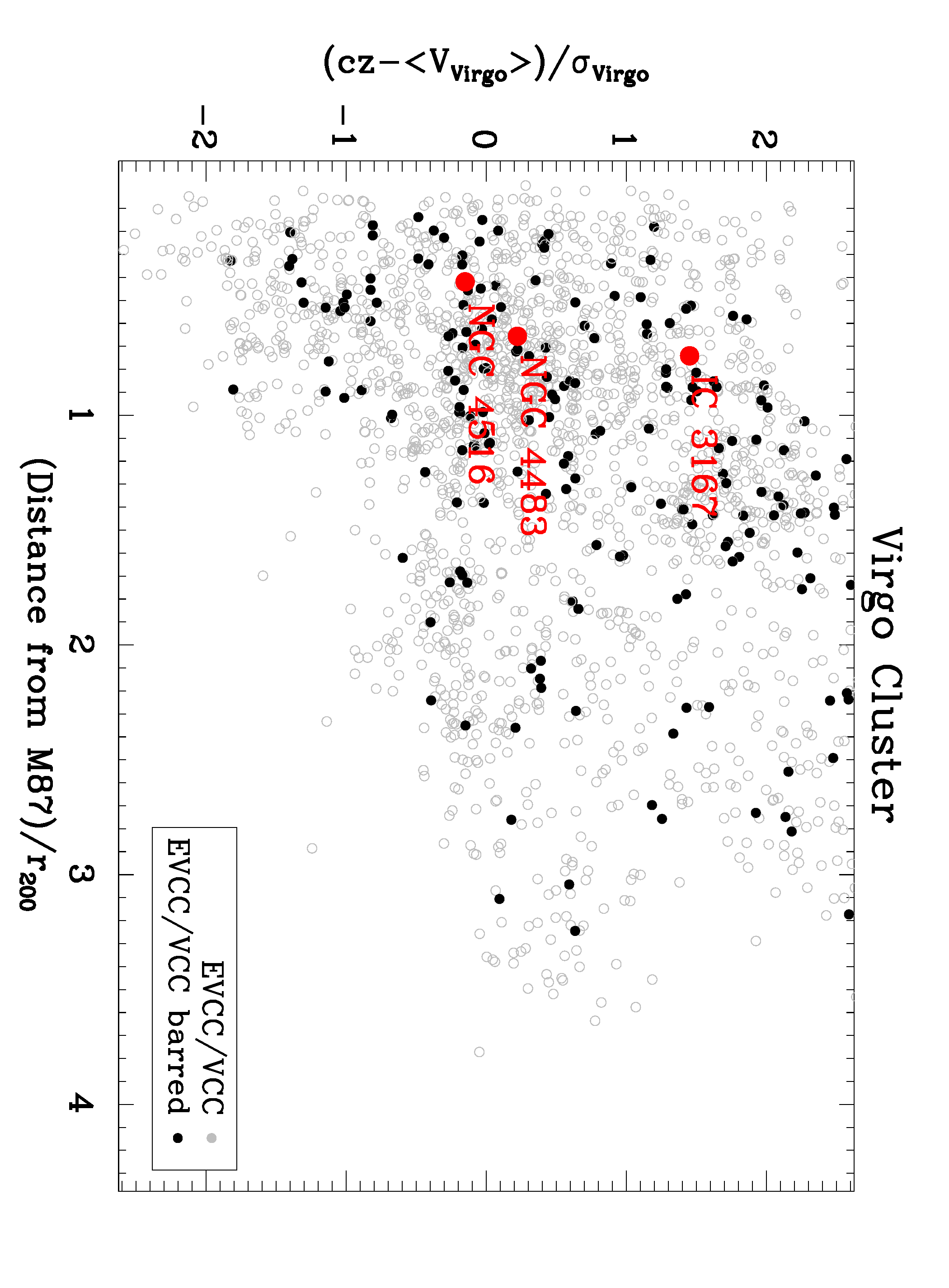}
    \caption{Phase-space plot of the Virgo Cluster. The cluster heliocentric velocity $V_{\rm Virgo}=1064$~km~s$^{-1}$ and velocity dispersion $\sigma_{\rm Virgo}=699$~km~s$^{-1}$ are taken from Binggeli et al. (1985) and virial radius $r_{200}=1550$~kpc from Yoon et al. (2017). The red circles mark the location of our dwarf barred galaxies. }
    \label{fig:galaxies_virgo_phase}
\end{figure}

Simulations suggest that bars are slow when the bar instability is induced by an interaction \citep{MartinezValpuesta2017,Lokas2018}. In this case, the rotation rate remains larger than 1.4 for many Gyr after the bar formation. Dwarf galaxies are more sensitive to modifications induced by interactions given their shallow potential well. Moreover, they are common in dense environments \citep{Binggeli1985}, making them the perfect candidates to host slowly rotating bars. \cite{Cuomo2022} suggested that the lopsided bar hosted in the dwarf galaxy IC~3167 could have been induced by the interaction with the Virgo Cluster, given that the galaxy is part of a group of galaxies falling in the cluster (Fig.~\ref{fig:galaxies_virgo_phase}). Indeed, an interaction explains both the slow rotation rate and the lopsided shape of the bar \citep{Lokas2021}. NGC~4483 and NGC~4516 are located in the central virialised region of the cluster, residing in region of intermediate projected surface density (Fig.~\ref{fig:galaxies_virgo_phase}). They may have suffered interactions with other galaxies and with the cluster itself. NGC~4516 hosts a bar with a high probability of being slowly rotating: bar formation driven by interaction in this dense environment is a plausible alternative explanation for the derived properties of this bar \citep{byrd1990,Valluri1993,Mastropietro2005}.

\subsubsection{Evolutionary effects}

In massive galaxies, longer and stronger bars tend to rotate with smaller \omegabar\ (see Fig. 4 from \citealt{Cuomo2020}). The observed \omegabar-$M_r$ anti-correlation is driven by secular evolution (see \citealt{Cuomo2020} and \citealt{Garma-Oehmichen2022} for a discussion). On the contrary, dwarf galaxies in our sample present a \omegabar-$M_r$ correlation. Given these galaxies are associated with low values of the circular velocities, the observed correlation is probably driven by the Tully-Fisher relation (Fig.~\ref{fig:tully_fisher}).

Finally, the bars analysed here turned out to be large with respect to the discs hosting them (with $R_{\rm bar}/h>1$). The corotation radii are even larger (with $R_{\rm cor}/h>1$, Fig.~\ref{fig:disc_param}). On the contrary, most of the massive galaxies with measurements of disc scale-lengths have both bar and corotation radii $<(1.0-1.5)h$. Simulated bars show that they evolve by redistributing the angular momentum in galaxies, evolving towards the slow regime \citep[see e.g.][]{Debattista2000,athanassoula2003,Athanassoula2013}. Using $N$-body simulations, \cite{Aguerri2023} showed that a simulated bar is expected to increase both its length and corotation radius with respect to the extension of the disc, being located in the same region of the $R_{\rm bar}/h-R_{\rm cor}/h$ plot as the bars in dwarf galaxies (see Fig. 7 from \citealt{Aguerri2023}). In this context, slow bars, which are also long with respect to the disc scale-length, should be more dynamically evolved with respect to fast and short bars. This is a plausible explaination of the properties of the bars in the dwarf galaxies we study.

\section{Summary and Conclusions}
\label{sec:conclusion}

In this paper we present the characterisation of the bars of the dwarf galaxies NGC~4483 and NGC~4516. By considering IC~3167 too, we discuss the properties of three dwarf galaxies with stellar masses between 1.2 and $7.5\times 10^9~\msun$, which are in different location within the Virgo Cluster. 

We perform a photometric analysis of NGC~4483 and NGC~4516 using the $i$- and $g$-band images from the NGVS to derive the disc properties and identify the bars. Moreover, we apply several photometric methods to derive bar length and strength. 

We measure the stellar kinematic maps of the LOS velocity and velocity dispersion using dedicated MUSE observations and derive the galaxy circular velocity applying a simple dynamical modelling based on the asymmetric drift correction. We then obtain the bar pattern speed applying the TW method using two different wavelength ranges (centred on the Mg I and CaT triplets, respectively) for the MUSE data, showing the results are consistent between each other. 

Given the bar length, pattern speed, and galaxy circular velocity, we then calculate the bar rotation rate. The bar in NGC~4483 and NGC~4516 have bar rotation rates in the slow regime, but we cannot exclude the fast rotation regime for NGC~4483. We discuss our findings including in the sample the lopsided bar in IC~3167. It was originally included in the observational campaign, but it is already analysed by \cite{Cuomo2022}. We compare the properties of these three dwarf barred galaxies with those of the massive barred galaxies from literature. 

In the following, we enumerate our main findings:

\begin{enumerate}
    \item Bars in dwarf galaxies are shorter, weaker, and slower (both in terms of the bar pattern speed and rotation rate) with respect to their massive counterparts. This may be in contrast with the expected secular evolution of bars, where the structures are born short, weak, and with large bar pattern speed and evolve towards long, strong, and with small bar pattern speed. This could be due to a different formation scenario of these bars. IC~3167 and NGC~4516 have bar rotation rates in the slow regime, in agreement with a formation scenario driven by interactions. 
    \item We found that the bar rotation rates of dwarf galaxies are larger than in more massive galaxies (i.e. slower rotating bars). Slow bars are expected to be associated with centrally-concentrated and massive DM halos, which slow down the bars through dynamical friction. Despite dwarf galaxies being the perfect candidates to host slowly-rotating bars, testing this scenario would require a complete dynamical modelling to derive the DM content within the bar region. 
    \item A weak correlation between the galaxy stellar mass and bar rotation rate has been highlighted. However, the number of dwarf galaxies explored so far with the TW method is still too small to drive strong conclusions. 
    \item Our analysis shows that the application of the TW method can be pushed towards the galaxy low-mass regime when a careful selection of the targets is performed and dedicated MUSE data are analysed. This will be pursued with forthcoming observational campaigns.
\end{enumerate}

\section*{Acknowledgements}
We are grateful to the anonymous referee, whose comments and suggestions have helped to greatly improve this manuscript.
VC acknowledges the support provided by ANID through 2022 FONDECYT postdoctoral research grant no. 3220206 and thanks Instituto de Astrof\'isica de Canarias and Università degli Studi di Padova for hospitality during the preparation of this paper. LM acknowledged the support from the concurso fondo ALMA 2021, grant no. ASTRO21-0007. VC and DG acknowledge the support from Comité Mixto ESO-Gobierno de Chile Grant no. ORP060/19. CB, EMC, AP and SZ are founded by MIUR grant PRIN 2017 20173ML3WW-001 and by the Padua University grant DOR2019-2021. AdLC acknowledges support from Ministerio de Ciencia e Innovación through the Spanish State Agency (MCIN/AEI) and from the European Regional Development Fund (ERDF) under grant CoBEARD (reference PID2021-128131NB-I00), and under Severo Ochoa Centres of Excellence Programme 2020-2023 (CEX2019-000920-S). SZ is supported by the Ministry of Science and Innovation of Spain, projects PID2019-107408GB-C43 (ESTALLIDOS) and PID2020-119342GB-I00, by the Government of the Canary Islands through EU FEDER funding, project PID2021010077. YHL acknowledges the support by the Basic Science Research Program through the National Research Foundation of Korea (NRF) funded by the Ministry of Education (RS-2023-00249435) and the Korea Astronomy and Space Science Institute (KASI) grant funded by the Korean government (MSIT; No. 2023-1-860-02, International Optical Observatory Project). The authors would like to extend appreciation to Narae Hwang for the insightful discussions.

\section*{Data Availability}

The derived data included in this article will be shared on request to VC.



\bibliographystyle{mnras}
\bibliography{biblio} 





\bsp	
\label{lastpage}
\end{document}